\documentclass[12pt]{article}
\usepackage{graphicx}
\usepackage{lscape}
\bibstyle{plain}

\newcommand{\be}{\begin{equation}}
\newcommand{\en}{\end{equation}}

\newcommand{\PP}{{\mathord{I\kern -.33em P}}}
\newcommand{\EE}{{\mathord{I\kern -.33em E}}}
\newcommand{\RR}{{\mathord{I\kern -.33em R}}}

\newcommand{\ea}{\end{eqnarray}}
\newcommand{\ba}{\begin{eqneduardarray}}
\newcommand{\ean}{\end{eqnarray*}}
\newcommand{\ban}{\begin{eqnarray*}}

\begin{document}

\title{Applying hedging strategies to estimate model risk and provision calculation.}

\author{Alberto Elices\thanks{Head of Equity Model Validation, Risk Methodology, Divisi\'{o}n General de Riesgos,  Santander, Ciudad Financiera Santander, Avda. Cantabria s/n, 28660 Boadilla del Monte, Spain, {\em aelices@gruposantander.com}.} \and Eduard Gim\'{e}nez \thanks{Head of Model Development Group, Front Office, CaixaBank, Av. Diagonal, 621-629 T.I. P13, Barcelona, Spain {\em eduard.gimenez.f@lacaixa.es}.}}
\date{\today}
\maketitle

\bigskip

\begin{abstract}
This paper introduces a relative model risk measure of a product priced with a given model, with respect to another reference model for which the market is assumed to be driven. This measure allows comparing products valued with different models (pricing hypothesis) under a homogeneous framework which allows concluding which model is the closest to the reference. The relative model risk measure is defined as the expected shortfall of the hedging strategy at a given time horizon for a chosen significance level. The reference model has been chosen to be Heston calibrated to market for a given time horizon (this reference model should be chosen to be a market proxy). The method is applied to estimate and compare this relative model risk measure under volga-vanna and Black-Scholes models for double-no-touch options and a portfolio of forward fader options.
\end{abstract}

\section{Introduction}
\label{sec:Introduction}

The optimistic situation during the years before the crisis in the second half of 2008 had trended from simple, marked to marked liquid products to progressively more complex and illiquid products which were marked to model. Although the degree of exoticity of products increased, they were still priced under considerably simple modeling assumptions. This led to a lot of reasonable but already limited variations of pricing models which in the very end were priced under refined assumptions of simple models such as Black-Scholes, Hull-White, Gaussian copula and so on. Wrong use of these simple models was one of the reasons which led to the 2008 crisis. After the crisis the trend has evolved toward less exotic and more liquid products valued under more complex pricing assumptions. However, there is still a big concern about model risk and an increasing trend toward provision or reserve calculation as an internal policy of the financial institution.

Markets such as foreign exchange continue trading liquidly first generation barrier products (barrier call/put, touch and no-touch options) which still exhibit a considerable model risk uncertainty. This means that although products are getting simpler, model risk estimation will be from now on a big concern anyway. Before the crisis, this first generation of FX exotic products were valued with the volga-vanna (VV) model (see \cite{Lipton2002}, \cite{Castagna2007}, \cite{Castagna2007b} and \cite{Wystup2003}). This is a heuristic analytical model based on the Black Scholes price with a smile correction given by the price difference of a portfolio of three vanilla options valued with smile and at-the-money volatilities. This portfolio of vanilla options is calculated to match the vega ($\frac{\partial{P}}{\partial{\sigma}}$), vanna ($\frac{\partial{P}^2}{\partial{\sigma}\partial{S}}$) and volga ($\frac{\partial{P}^2}{\partial{\sigma}^2}$) of the barrier Black-Scholes price ($P$) with at-the-money volatility. It has been justified (see \cite{Castagna2007b}) that this model provides accurate results for non-path-dependent options (e.g. quanto options). However, nothing could be concluded for path-dependent options such as barriers. 

A provision calculation philosophy should cover the expected hedging loss plus some of the uncertainty of that loss. Some characteristics of a good provision policy is the fact that they should be transparent and easy to compute (so that both Front Office and Risk department can calculate them easily). They should be dynamic (change through time) and stable with a smooth decreasing evolution as expiry approaches (the closer to expiry the less uncertainty and therefore the less model risk provision). From a management perspective, provisions should balance risk mitigation with business limitation. This means that they can provide the means to approve marketing derivative products to customers using limited models with controllable risk. For instance, if a deal can be valued with a slow but accurate model but the number of deals is huge, product marketing could be approved with a simpler less accurate model if the appropriate provision is applied. In addition, provisions introduce an incentive to improve models. A less accurate model may be charged a high provision so that when it gets improved, the provision is reduced\footnote{Provisions may only delay the realized profit and loss of traders. They are subtracted from the expected earnings reported by the trader on a given deal to justify the objectives of the year. When the provision gets reduced through time, those earnings are released and the trader can report them for the objectives of the following years.}.

Model Risk can arise out of many different sources of uncertainty: unrealistic model hypothesis, inaccurate calibration, different calibration strategies, different sets of calibration products or externally input parameters which are not calibrated to market (e.g. correlations, dividends, mean reversion). Therefore, model risk could be estimated on a deal-by-deal basis by calculating sets of prices varying these sources of uncertainty. For example, a set of prices can be calculated varying one or more of these sources of uncertainty: model hypothesis (e.g. change mean reversion, correlations or evaluate same deal with different models), calibration sets, calibration strategies or externally-input parameters not calibrated to market. This set of prices gives an interval which could be interpreted as a bid-offer spread (this bid-offer spread would be theoretically composed by the set of arbitrage-free prices according to the arbitrage free measures given by all possible underlying hypotheses or ``models''). The absolute value of the difference between the highest price (offer side) and the lowest price (bid side) would give an estimate for model risk for a single deal. The key question now would be which parameters to change and how much they should be changed. In addition, this strategy is usually carried out on a deal-by-deal basis and portfolio diversification is not considered. Furthermore, each product may have its own way of estimating model risk. In this context, the provision would be a fraction or multiplier of this model risk quantity. Through this approach, the senior management can clearly define their risk profile.

This ``absolute'' way of measuring model risk is formalized and analyzed in a systematic way by \cite{Cont2006}. In this reference, a given product is valued under different arbitrage-free pricing measures which must satisfy certain properties. Each measure gives a different price and all the prices are confined in an interval (a bid-offer spread). The model risk measure is then defined as the difference between the offer and bid prices. This can be regarded as an ``absolute'' model risk measure, because any pricing hypothesis can virtually be considered and the interval can get progressively broader as long as more pricing measures are incorporated. In practice, only a limited number of measures can be considered (e.g. changing unobservable market parameters, calibration sets or hypotheses of the evolution of the underlying risk factors). This limited number of pricing measures allow estimating model risk. The portfolio effect can be considered when the whole portfolio is priced with each measure and model risk is calculated from the bid-offer spread obtained for the whole portfolio.

This paper does not address this ``absolute'' model risk approach. Instead, it is concerned with a relative model risk measure of a particular product priced with a given model with respect to a reference model under which the market dynamics are assumed to be driven. The relative measure is defined by the expected shortfall at a given significance level of the hedging strategy of the product priced according to its model, assuming that the market is driven by the reference model. The advantage of the expected shortfall measure is the fact that it is coherent (see \cite{Artzner1999}). This means that it satisfies a number of properties which provide a framework to deal with model risk in a systematic way. This relative measure provides a homogeneous framework to compare  products priced with different models under the same reference. The products chosen are double-no-touch and fader options, the models under analysis are volga-vanna and Black-Scholes and the reference model is Heston. The ideal situation would be that the reference model is chosen to be a market proxy. The original motivation of this work was to have a reasonable criterion to decide from a list of models which one was the most appropriate to price and manage a product. If the risk department is worried because many double-no-touch deals are closed what would be your answer if your boss asks you: what would you use to manage these positions? A heuristic model such as the volga-vanna or the well-known Black-Scholes model but without skew pricing hypothesis? Getting back to basics, simulating the hedging strategy seemed a reasonable way to answer this question.

Considering hedging strategies to contrast valuation models is not new (see for instance \cite{Kurpiel1998}, \cite{Carr2007} and \cite{Carr2002}). In this context, the main goal of this paper is to estimate relative model risk in a way which could be consistent and robust, general purpose and applicable to a portfolio of deals. The approach followed by this paper is to simulate the hedging strategy. Therefore, only analytical models can be considered in practice as the computational cost is very high. That's why this study was carried out with the help of a grid of computers. Two assumptions are considered for the hedging strategy. The first one assumes that the market volatility surface is driven by Heston's model (see \cite{Heston1993}) calibrated to market at a given time horizon. The second one is the fact that hedging inaccuracies from using actual sensitivities provided by the pricing model or using several vanillas to hedge the movements of the volatility surface are eliminated. This is possible because market dynamics are known and the hedging ratios are calculated to hedge the two known factor which drive the market. This methodology can be applied to a single deal or a portfolio of them. Every model is evaluated under the same framework and therefore, different models and products can be quantitatively compared with each other. A reasonable provision would be equal to this relative model risk measure. This method will be applied to a single deal (double-no-touch option) and a portfolio of fader options with the aid of a grid of computers. The first one will be valued under Black Scholes and volga-vanna hypothesis and the second one only using volga-vanna.

Under the Black-Scholes \cite{Black1973} framework, i.e. the asset follows a lognormal distribution on the real world measure with drift $\mu$, there exists a clear mapping between the real world probability $\mathcal{P}$ and the risk neutral probability $\mathcal{Q}$ used for pricing and hedging. This is also true under Dupire's framework \cite{Dupire1994}. Apart from these two models it is difficult to find other with such direct relationship.  For example, Merton \cite{Merton1993} jump diffusion model assumes that jumps are part of the firm idiosyncratic risk. Since the market portfolio cancels all firm idiosyncratic risk by diversification, according to Capital Asset pricing model \cite{Sharpe1964}, no risk premium should be taken into account. Heston \cite{Heston1993} model needs to assume that investors have a very particular utility function to derive the partial differential equation of the option price and find a correspondance between pricing measures $\mathcal{P}$ and $\mathcal{Q}$. Beyond these ideal cases authors have become more and more aware that a perfect hedging might not be possible either because the market is incomplete (a unique risk neutral probability does not exist) or because for practical reasons traders might want to use a simpler and faster model for pricing and hedging. The former case has been extensively studied by Scheweizer \cite{Schweizer1992}, contributing to the mean-variance hedging technique and Zhou \cite{Zhou2007}, embedding the problem into a ``Markowitz'' type problem. The latter case has been treated by Corielli \cite{Corielli2006} using $L^p$ energy inequities for parabolic equations which allow estimating expected pricing and hedging errors between the ``true'' model and the one used by the trading desks.  Our work could be regarded as a practitioner approach to this latter type of problem. Although Heston's model has been used as ``real world probability'' it could be argued that as long as the Risk department is able to provide a coherent evolution of the underlying and the vanilla options, the methodology would still hold. Nevertheless, the presented methodology is not directly meant to solve the mean-variance hedging problem rather than to study for a given model developed by a Front Office team, how well will it behave as market conditions are stressed.

The paper starts with the formulation of the hedging strategy in section \ref{sec:Formulation}. Section \ref{sec:DNT} applies the methodology to double-no-touch options valued with Black Scholes and volga-vanna models. Section \ref{sec:ModelRiskMeasure} defines the relative model risk measure which will allow comparing different models in the following sections. Section \ref{sec:DNT} applies the methodology to double-no-touch options valued with Black-Scholes and volga-vanna models. It is concluded that double-no-touch products have high model risk and hedging with volga-vanna is better than Black-Scholes. Section \ref{sec:Faders} applies the methodology to a portfolio of forward fader options using only the volga-vanna model. It is concluded that model risk is irrelevant for this product. Finally, section \ref{sec:Conclusions} ends with some conclusions.


\section{Formulation of the hedging strategy}
\label{sec:Formulation}

This section presents the formulation of the hedging strategy which will be used to compare different products evaluated with different models. Subsection \ref{sec:MarketSimulation} shows the hypothesis assumed for the market and how to calculate the hedge ratios. Subsection \ref{sec:ImplementationHedging} explains the implementation of the hedging strategy through an intuitive derivation. The rigorous derivation of the hedging strategy is presented in subsection \ref{sec:RigorousDerivationHedging} justifying why model risk leads to a drift in the profit or loss of the hedging strategy and quantifying it on average.

\subsection{Calculation of hedge ratios and market simulation}
\label{sec:MarketSimulation}

The main and hardest assumption of this hedging strategy is that the market is driven by Heston's process given by equation (\ref{eq:MarketProcess}), where $S_t$ is the underlying spot price at time $t$, $r_t^d$ and $r_t^f$ are the time-dependent domestic and foreign interest rates, $v_t$ is the variance process, $\theta$ is the long-term variance, $dW_t$ and $dV_t$ are correlated brownian motions with correlation $\rho$ ($d\left\langle {W_t ,V_t } \right\rangle  = \rho dt$) and $\eta$ is the volatility of the variance process. 

\begin{equation}
\begin{array}{l}
\frac{dS_t}{S_t}  = \left( {r_t^d  - r_t^f } \right)dt + \sqrt {v_t } dW_t  \\ 
 dv_t  = \kappa \left( {\theta  - v_t } \right)dt + \eta \sqrt {v_t } dV_t  \\ 
 \end{array}
  \label{eq:MarketProcess}
\end{equation}

This model has two risk factors: the evolution of the underlying $S_t$ and the variance $v_t$. Therefore, on a given date, the set of model parameters ($\kappa$, $\theta$, $\eta$ and $\rho$) and these two factors completely define the market. This means that the spot and forward volatility surface (out of vanilla and forward start call and put options) are completely defined (see \cite{Elices2009} and \cite{Elices2007} for a semi-analytical implementation of Heston's pricing for spot and forward start vanilla options). This assumption that the market is driven by Heston's process might not be realistic (even if it is calibrated to market for a given horizon). However, it will be seen in the following sections that in spite of this fact, this hypothesis provides a framework in which different models can be compared.

Under this assumption, the evolution of the market is obtained by simulating the factors $S_t$ and $v_t$ at each point of time using a set of model parameter which were obtained calibrating the model to real market on a particular date. From these factors, the implied volatility surface is calculated for each day and path simulated by this hypothetical market. The hedging strategy is built in order to neutralize the uncertainty of these two factors. The hedging products used are the underlying and an at-the-money 6 month vanilla call option. Thus, the sensitivities of the premium of the product being hedged and the hedging products (the 6 month vanilla option) are calculated with respect to $S_t$ and $v_t$ as indicated by equation (\ref{eq:GreekDefinition}), where $\delta_S = S_t \cdot 10^{-4}$ and $\delta_v = (10^{-3})^2$.

\begin{equation}
\begin{array}{*{20}c}
   {\Delta  = \frac{{\partial P}}{{\partial S_t }} = \frac{{P(S_t  + \delta _S ) - P(S_t  - \delta _S )}}{{2\delta _S }}} & {} & {\vartheta  = \frac{{\partial P}}{{\partial v_t }} = \frac{{P(v_t  + \delta _v ) - P(v_t )}}{{\delta _v }}}  \\
\end{array}
  \label{eq:GreekDefinition}
\end{equation}

The premium $P(v_t+\delta_{v})$ is calculated by tweaking $v_t$, rebuilding the whole volatility surface with an initial variance $v_t + \delta_v$ and repricing the product with the rebuilt volatility surface (for the 6 month vanilla it is only necessary to change the initial variance leaving Heston's parameters constant). The same should be done for $P(S_t + \delta_S)$. However, as the FX volatility surface is expressed in terms of delta (rather than strike), the volatility surface does not change when $S_t$ is varied\footnote{As the delta points of the volatility surface are fixed, the moneyness ratio (strike over forward) of those points is kept constant. Therefore, as the pricing formulas for Heston's model depend on moneyness, the prices will not change and neither will do the implied volatilities.}. It is important to stress that $\Delta$ is the standard FX delta (moving the spot without changing the implied volatility surface). However, it would not be the standard delta in an equity context (the surface should be rebuilt changing $S_t$). Concerning $\vartheta$, it is clearly not the standard vega (premium sensitivity under a parallel movement of the volatility surface). These sensitivities are ``external'' in the sense that they are not calculated by the pricing model itself but by the hedging algorithm to properly hedge the two factors driving the hypothetical market. In this context, the sensitivities given by the model are ignored, only the premium is used.

\begin{equation}
\vartheta  = \frac{{dP}}{{dv_t }} = \sum\limits_{i = 1}^N {\frac{{\partial P}}{{\partial \sigma _i }}\frac{{\partial \sigma _i }}{{\partial v_t }}} 
  \label{eq:vega_model}
\end{equation}

If vega sensitivities provided by the model were used to compute $\vartheta$, equation (\ref{eq:vega_model}) should be used instead of equation (\ref{eq:GreekDefinition}), where $\frac{{\partial P}}{{\partial \sigma _i }}$ is the actual vega returned by the model at different maturity and delta buckets and $\frac{{\partial \sigma _i }}{{\partial v_t }}$ is the sensitivity of Heston's implied volatility of each bucket with respect to the change of the initial variance $v_t$. Under Heston's model, implied volatility movements with respect to $v_t$ have only term structure (all delta buckets move parallel to each other). This means that to hedge the risk factor $v_t$, only the term structure of vega returned by the model would be needed (the sum of equation (\ref{eq:vega_model}) would be only across maturities). If the market were represented by a mixed volatility model (see \cite{Ren2007}), the movements of the implied volatility surface would depend on both the spot level and the initial variance and the implied volatility sensitivities with respect to the initial variance would not be equal across delta buckets. In addition, a vanna contribution $\frac{{\partial^2 P }}{{\partial S_t\partial v_t }}$ should be added.

The reason to calculate $\vartheta$ according to equation (\ref{eq:GreekDefinition}) instead of equation (\ref{eq:vega_model}) is to avoid the hedging noise of the prediction error of the premium change obtained out of model sensitivities for the given movement of the implied volatility surface. However, comparing the final distribution of the hedging cost when $\vartheta$ is calculated according to equations (\ref{eq:GreekDefinition}) and (\ref{eq:vega_model}) would provide a quantitative measure of the quality of the sensitivities provided by the model and their capability of predicting premium change (this would be very useful for model validation purposes).

See that a regular trader would only know $\frac{{\partial P}}{{\partial \sigma _i }}$ (the regular vega sensitivity provided by the model used to manage the position) but not $\frac{{\partial \sigma _i }}{{\partial v_t }}$ because the dynamics of the market are unknown for the trader. However, a regular trader would not use a single option (e.g. 6 month at-the-money) as the study presented here, but most likely three vanillas (to account for the level, steepness and convexity of the volatility surface) for a number of maturities. In the study presented in this paper, it is possible to hedge all movements of the volatility surface with a single vanilla because the dynamics of the market are known and therefore the movements of the volatility surface with respect to the market driving factors ($S_t$ and $v_t$). In a more realistic situation, the dynamics of the market would be unknown but several vanilla options would be used for hedging leading to a similar result. The more realistic approach is not followed here because the hedging would be worse and the purpose of this paper is to provide an estimation of the model risk premium, filtering out the imperfections of hedging with several vanillas and using the sensitivities provided by the model. Comparing model risk when these imperfections are not filtered out is also very interesting but will be left for a future research.

As the foreign exchange implied volatility surface is expressed in terms of delta (and not strike), it is build through an iterative method. This method starts with an initial term structure of implied volatility given by the solution of the deterministic part of the variance process ($dv_t  = \kappa \left( {\theta  - v_t } \right)dt$). The FX volatility surface conventions are converted into the simple Black Scholes setting (spot delta, premium not included in delta, ATM straddle, call and put on foreign currency). Strikes are calculated out of the surface delta buckets and the initial term structure of implied volatility. From the strikes, vanilla call option prices are calculated according to Heston's model with the given set of parameters and their implied volatilities are calculated thereafter. Black Scholes call deltas are then calculated out of the strikes and implied volatilities. They will not coincide with the delta buckets of the surface expressed as call option deltas. Therefore, the implied volatilities of the delta buckets are interpolated from the just calculated delta versus implied volatility points and the process starts again. Strikes are calculated with the new volatilities, then prices, implied volatilities and deltas which get compared with those of the buckets. The process repeats until the deltas obtained in the iteration are close enough to the delta buckets of the surface.

\subsection{Implementation of hedging strategy}
\label{sec:ImplementationHedging}

This section presents the implementation of the hedging strategy using an intuitive derivation. The rigorous justification for continuous time will be provided in subsection \ref{sec:RigorousDerivationHedging} to interpret the findings of sections \ref{sec:DNT} and \ref{sec:Faders}. Consider now a portfolio $\Pi^{Tot}_t$ composed by the portfolio of options to hedge $\Pi_t$ (this is not cash but positions on option contracts) valued with the model under test and the hedging portfolio $\Pi_t^{Hedge}$ as indicated by equation (\ref{eq:PITotal}). Equation (\ref{eq:PIHedgeBf}) shows that the hedging portfolio is composed by a cash amount $B_t$ in domestic currency which will get funded, an amount of underlying $\alpha_t$ and $\beta_t$ units of the 6 month vanilla call option whose price is denoted by $C_t$. See that the underlying security is not just $S_t$ but $S_t B_t^f$, where $B_t^f$ is the process for the evolution of the foreign bank account given by equation (\ref{eq:Bt_f}) and $r_t^f$ is the foreign risk free rate (an equivalent ``bank account'' could be defined for dividend payments). This underlying is the real traded security (the foreign bank account $B_t^f$ expressed in domestic units) and that is why is is used as the underlying security.

\begin{equation}
\Pi _t^{Tot}  = \Pi _t  + \Pi _t^{Hedge} 
  \label{eq:PITotal}
\end{equation}

\begin{equation}
   \Pi _t^{Hedge}  = B_t  + \alpha _t  \cdot S_t B^f_t + \beta _t  \cdot C_t
  \label{eq:PIHedgeBf}
\end{equation}

\begin{equation}
{dB_t^f  = r_t^f B_t^f dt}
  \label{eq:Bt_f}
\end{equation}

Consider a set of instants of time $[t_0 ,t_1 , \cdots ,t_N ]$ up to a final horizon $t_N$ (intervals are usually daily or even more frequent) where the hedging ratios $\alpha_{t_i}$ and $\beta_{t_i}$ will change.


Assuming no dependence of the product and the hedging porfolio with respect to time (time is frozen) the condition to derive $\alpha_t$ and $\beta_t$ would be given by equation (\ref{eq:DPITot_eq0}). Namely, there is neither gain nor loss in each time interval.


\begin{equation}
\Delta\Pi _t  + \Delta\Pi _t^{Hedge}  = 0
  \label{eq:DPITot_eq0}
\end{equation}

Equations (\ref{eq:DPI}) and (\ref{eq:DPIHedge}) show an approximation of the change of $\Pi_t$ and $\Pi^{Hedge}_t$, where $\Delta^{\Pi}_t$ and $\Delta^{C}_t$ are the deltas of the portfolio $\Pi_t$ and $C_t$, and $\vartheta^{\Pi}_t$ and $\vartheta^{C}_t$ are their sensitivities with respect to the variance process $v_t$. These sensitivities are calculated according to equation (\ref{eq:GreekDefinition}). Replacing equations (\ref{eq:DPI}) and (\ref{eq:DPIHedge}) in (\ref{eq:DPITot_eq0}) yields the values of $\alpha_t$ and $\beta_t$ given by equation (\ref{eq:AlphaBeta}).

\begin{equation}
\Delta \Pi _t  = \Delta _t^{\Pi}  \Delta S  + \vartheta _t^\Pi  \Delta v 
  \label{eq:DPI}
\end{equation}

\begin{equation}
\begin{array}{l}
 \Delta \Pi _t^{Hedge}  = \alpha _t B_t^f \Delta S  + \beta _t \left( {\Delta _t^C \Delta S  + \vartheta _t^C \Delta v } \right) \\ 
 \;\;\;\;\;\;\;\;\;\;\;\;\;\;
   = \left( {\alpha _t B_t^f + \beta _t \Delta _t^C } \right) \Delta S  + \beta _t \vartheta_t^C \Delta v  \\ 
\end{array}
  \label{eq:DPIHedge}
\end{equation}

\begin{equation}
\begin{array}{*{20}c}
   {\alpha _t =  - \frac{\Delta _t^\Pi   + \beta _t \Delta _t^C }{B_t^f}} & {} & {\beta _t  =  - \frac{{\vartheta _t^\Pi  }}{{\vartheta _t^C }}}  \\
\end{array}
  \label{eq:AlphaBeta}
\end{equation}

See that the sensitivities $\Delta_t^{\Pi}$ and $\vartheta_t^{\Pi}$ of equation (\ref{eq:DPI}) are calculated using the model under analysis, whereas the sensitivities $\Delta_t^C$ and $\vartheta_t^C$ are calculated with a premium according to the model assumed for the market. Equation (\ref{eq:HHedge}) shows the value of the hedging position $(\alpha_{t_i},\beta_{t_i})$ at time $t_j$ with hedge ratios calculated at time $t_i$.

\begin{equation}
{\rm H}_{t_j }^{t_i }  = \alpha _{t_i } \cdot S_{t_j } B_{t_j}^f + \beta _{t_i }  \cdot C_{t_j } 
  \label{eq:HHedge}
\end{equation}

The steps of the hedging strategy start with a given portfolio of deals or a new deal which is bought or sold. At a given point of time, the hedging strategy starts with the price $\Pi_{t_0}$ of a deal or portfolio. This price can be calculated with a range of models (local volatility, volga-vanna, Black Scholes, Heston and so on). This price is positive for a net long position and negative for a short position. The hedging portfolio would then start with a cash position,  $B_{t_0} = -\Pi_{t_0}$, and no position in underlying or options ($\alpha_{t_0} = 0$ and $\beta_{t_0} = 0$). If $\Pi_{t_0}$ is positive, this means that a long position is held and it has to be financed by borrowing money at the inter-bank prevailing rate\footnote{Although funding issues are beyond the scope of this paper, with a collateral agreement, this amount of money would be deposited by the counterparty into a collateral account which should be paid on a daily basis at the prevailing collateral rate (e.g. EONIA). This note is simply for information purposes. The hedging simulations carried out in this paper do not consider funding issues.}. If $\Pi_{t_0}$ is negative, a net short position would be considered instead and the equivalent amount of money would be received and it could be lent at the prevailing inter-bank lending rate\footnote{With a collateral agreement, the amount $\Pi_{t_0}$ should be deposited into the collateral account of the counterparty and the collateral prevailing rate (e.g. EONIA) would be received on a daily basis.}.

Ignoring transaction costs, the implemented hedging routine process would be given by the following steps:

\begin{itemize}
	\item Sell the hedging position $(\alpha_t, \beta_t)$ bought in the previous instant (the first time there is no position).
	\item Buy a new hedging position with the updated hedge ratios\footnote{In practice, the hedging portfolio is not sold and bought again, but just rebalanced.}.
	\item Between selling and buying the hedging portfolio, management of coupon payments\footnote{A Discrete dividend payment would not have impact as the income to the cash account would be compensated by a down jump move in the underlying spot.}, strike fixings, barrier touching, cancellations, expiration of options and so on should be carried out (incoming payments are positive in the cash account $B_t$ and outgoing ones are negative).
	\item Add interest earned (if cash amount is positive) or paid (when it is negative) on the cash amount in domestic currency and add foreign interest earned on the the underlying position in foreign currency.
\end{itemize}

The last step accounts the financing costs (depending whether the balance is positive or negative or there is a collateral agreement). In an equity context, continuous dividend payments should be deposited into the cash account, causing a falling drift of the underlying process.

\begin{equation}
\Pi _{t_0 }^{Hedge}  = \left( { - \Pi _{t_0 }  - {\rm H}_{t_0 }^{t_0 } } \right) + \alpha _{t_0 } S_{t_0 } B^f_{t_0}  + \beta _{t_0 } C_{t_0 } 
  \label{eq:PIHedge_t0}
\end{equation}

\begin{equation}
\Pi _{t_1 }^{Hedge}  = \left( { - \frac{\Pi _{t_0 }}{P^d_{t_0,t_1}}  - \frac{ {\rm H}_{t_0 }^{t_0 } }{P^d_{t_0,t_1}} + {\rm H}_{t_1 }^{t_0 }  - {\rm H}_{t_1 }^{t_1 } } \right) + \alpha _{t_1 } S_{t_1 } B^f_{t_1} + \beta _{t_1 } C_{t_1 } 
  \label{eq:PIHedge_t1}
\end{equation}

\begin{equation}
\begin{array}{l}
 \Pi _{t_N }^{Hedge}  = \left( { \frac{ -\Pi _{t_0 }}{P^d_{t_0,t_N}}  + \sum\limits_{i = 1}^N {\left( { \frac{{\rm H}_{t_i }^{t_{i - 1} }}{P^d_{t_i,t_N}}  - \frac{{\rm H}_{t_{i - 1} }^{t_{i - 1} }}{P^d_{t_{i-1},t_N}} } \right)}  - {\rm H}_{t_N }^{t_N } } \right) \\ 
 \;\;\;\;\;\;\;\;\;\;\;\;\;\;\;
 + \alpha _{t_N } S_{t_N } B^f_{t_N} + \beta _{t_N } C_{t_N }  \\ 
 \end{array}
  \label{eq:PIHedge_tN}
\end{equation}

Equation (\ref{eq:PIHedge_t0}) shows the resulting hedging portfolio on the first date (the amount in parenthesis is held in cash). See that the only thing which has been done is to buy the new hedge position $(\alpha_{t_0},\beta_{t_0})$ and its price, ${\rm H}_{t_0 }^{t_0 }$, has been subtracted from the cash account. Equation (\ref{eq:PIHedge_t1}) shows the hedging portfolio at $t_1$ which has been held one period and therefore the cash position gets accrued by $(P^d_{t_0,t_1})^{-1}$, where $P^d_{t_i,t_j}$ is the bond value on $t_i$ of a bond expiring on $t_j$. See that the hedging position is sold at the beginning of the second period with the positive income ${\rm H}_{t_1 }^{t_0 }$ of the hedging position of the previous instant $t_0$ (valued at $t_1$) and the negative outcome, ${\rm H}_{t_1 }^{t_1 }$, to buy the new hedging position $(\alpha_1, \beta_1)$. Finally, equation (\ref{eq:PIHedge_tN}) shows the hedging portfolio for a generic instant $t_N$. The summation term shows the cumulative profit or loss of holding the hedging position from one instant to the following. The negative term ${\rm H}_{t_N }^{t_N }$ is the cash outflow to buy the hedging position $(\alpha_{t_N}, \beta_{t_N})$ at time $t_N$.

The construction of the evolution of the hedging portfolio of equation (\ref{eq:PIHedge_tN}) helps understanding how the implementation is carried out in discrete time. Although the derivation of the hedging ratios is intuitive but not rigorous, section \ref{sec:RigorousDerivationHedging} shows that the hedge ratios are indeed those obtained in equation (\ref{eq:AlphaBeta}). In addition, this section will also show that when the hedge ratios are calculated with a model whose pricing assumptions do not reflect the market driving factors, a consistent profit or loss leakage will be seen on a daily basis in the cash account of the hedging portfolio.

The formerly described hedging strategy is computationally expensive even if the underlying portfolio of options is priced with analytical models. This is so because Front Office pricing models require the underlying spot, interest curve and a detailed volatility surface as inputs. Therefore, at each point in time it is necessary to calculate the complete volatility surface consistent to Heston's parameters for the hypothetical market from the simulated $S_t$ and $v_t$.

\begin{equation}
{\rm request  = }\left[ {\left. {\left( {S_{t_1 } ,v_{t_1 } } \right), \cdots ,\left( {S_{t_N } ,v_{t_N } } \right)} \right|\left( {Op_1  \cdots Op_m } \right)} \right]
  \label{eq:request}
\end{equation}

To perform this task a computer grid was set up. On the client side a set of paths ($S_t$, $v_t$) was generated. Each one of these paths together with the portfolio of options under management, $\Pi_t$, would become a request message as indicated by equation (\ref{eq:request}).

\begin{equation}
{\rm result  =  }\left[ {\left( {\Pi _{t_1 }^{Tot} ,\Pi _{t_1 } ,\Delta _{t_1 } ,\vartheta _{t_1 } } \right), \cdots ,\left( {\Pi _{t_N }^{Tot} ,\Pi _{t_N } ,\Delta _{t_N } ,\vartheta _{t_N } } \right)} \right]
  \label{eq:result}
\end{equation}

On the server side, each $(S_t, v_t)$ path (the calculation of the volatility surface together with the hedging strategy) is executed in a single server computer out of the total number of servers of the grid. The result message would contain the evolution of the profit and loss $\Pi^{Tot}_t$, the value of the whole portfolio $\Pi_t$, $\Delta_t$ and $\vartheta_t$ over time as showed by equation (\ref{eq:result}).

\subsection{Rigorous derivation of hedging strategy}
\label{sec:RigorousDerivationHedging}

Consider the change of value of the portfolio $\Pi^{Tot}_t$ given by equation (\ref{eq:dDiscPI}) which is composed by the options to hedge $\Pi_t$ and the hedging portfolio $\Pi^{Hedge}_t$ given by equation (\ref{eq:PIHedgeBf}).

\begin{equation}
d\Pi^{Tot}_t = d\Pi _t  + d\Pi _t^{Hedge}  = r_t^d \left( {\Pi _t  + \Pi _t^{Hedge} } \right)dt
  \label{eq:dDiscPI}
\end{equation}

When the whole randomness has been hedged out, the portfolio, $\Pi^{Tot}_t$ should earn the risk free rate (in this example the domestic rate $r^d_t$) as indicated by equation (\ref{eq:dDiscPI}).

\begin{equation}
\begin{array}{l}
 df = \frac{{\partial f}}{{\partial t}}dt + \frac{{\partial f}}{{\partial S_t }}dS_t  + \frac{{\partial f}}{{\partial v_t }}dv_t  +  \\ 
 \begin{array}{*{20}c}
   {} & {} & {}  \\
\end{array} + \frac{1}{2}\frac{{\partial ^2 f}}{{\partial S_t^2 }}d\left\langle {S_t ,S_t } \right\rangle  + \frac{1}{2}\frac{{\partial ^2 f}}{{\partial v_t^2 }}d\left\langle {v_t ,v_t } \right\rangle  + \frac{{\partial ^2 f}}{{\partial v_t \partial S_t }}d\left\langle {v_t ,S_t } \right\rangle  \\ 
 \end{array}
  \label{eq:df_Ito}
\end{equation}

\begin{equation}
\begin{array}{*{20}c}
   {d\left\langle {S_t ,S_t } \right\rangle  = S_t^2 v_t dt} & {d\left\langle {v_t ,v_t } \right\rangle  = \eta ^2 v_t dt} & {d\left\langle {v_t ,S_t } \right\rangle  = \rho \eta S_t v_t dt}  \\
\end{array}
  \label{eq:QuadraticVar}
\end{equation}

Equation (\ref{eq:df_Ito}) shows the dynamics of any pricing model, $f$, whose inputs are the spot price and the volatility surface given by the model assumed for the market (the function $f$ only depends on the factors $S_t$ and $v_t$). The dynamics are obtained by applying It\^{o}'s formula to the function $f$. The quadratic variations corresponding to the model assumed for the market are given by equation (\ref{eq:QuadraticVar}).

\begin{equation}
dC_t  = \mathcal{L}^{mkt}C_t dt + \Delta _t^C S_t \sqrt {v_t } dW_t  + \vartheta _t^C \eta \sqrt {v_t } dV_t 
  \label{eq:dCt}
\end{equation}

\begin{equation}
d\Pi _t  = \mathcal{L}^{mkt}\Pi _t dt + \Delta _t^\Pi  S_t \sqrt {v_t } dW_t  + \vartheta _t^\Pi  \eta \sqrt {v_t } dV_t 
  \label{eq:dPI}
\end{equation}

\begin{equation}
\begin{array}{l}
 \mathcal{L}^{mkt} = \frac{\partial }{{\partial t}} + \frac{1}{2}S_t^2 v_t \frac{{\partial ^2 }}{{\partial v_t^2 }} + \rho \eta S_t v_t \frac{{\partial ^2 }}{{\partial v_t \partial S_t }} + \frac{1}{2}\eta ^2 v_t \frac{{\partial ^2 }}{{\partial v_t^2 }} +  \\ 
 \begin{array}{*{20}c}
   {} & {} & {}  \\
\end{array} + \kappa \left( {\theta  - v_t } \right)\frac{\partial }{{\partial v_t }} + r_t^d S_t \frac{\partial }{{\partial S_t }} \\ 
 \end{array}
  \label{eq:Loperator}
\end{equation}

Equations (\ref{eq:dCt}) and (\ref{eq:dPI}) show the dynamics of the option to hedge, $\Pi_t$, and the 6 month call option used for hedging, $C_t$. These expressions are obtained replacing the quadratic variations (\ref{eq:QuadraticVar}) and the market dynamics (\ref{eq:MarketProcess}) in equation (\ref{eq:df_Ito}) for $f=\Pi_t$ and $f=C_t$. The coefficients $\Delta^C_t$, $\Delta^{\Pi}_t$, $\vartheta^C_t$ and $\vartheta^{\Pi}_t$ are obtained using equation (\ref{eq:GreekDefinition}) and $\mathcal{L}^{mkt}$ given by equation (\ref{eq:Loperator}) is the differential operator (infinitesimal generator) of Heston's process assumed for the market.

\begin{equation}
\begin{array}{l}
 d\Pi _t^{Hedge}  = dB_t  + \alpha _t d\left( {S_t B_t^f } \right) + \beta _t dC_t  =  \\ 
 \begin{array}{*{20}c}
   {} & {}  \\
\end{array} = dB_t  + \alpha _t \left( {B_t^f dS_t  + S_t dB_t^f } \right) + \beta _t dC_t  \\ 
 \begin{array}{*{20}c}
   {} & {}  \\
\end{array} = dB_t  + \alpha _t \left( {B_t^f \left[ {(r_t^d  - r_t^f )S_t dt + \sqrt {v_t } S_t dW_t } \right] + S_t dB_t^f } \right) + \beta _t dC_t  \\ 
 \begin{array}{*{20}c}
   {} & {}  \\
\end{array} = \left( {r_t^d B_t  + \alpha _t r_t^d S_t B_t^f } \right)dt + \alpha _t \sqrt {v_t } S_t B_t^f dW_t  + \beta _t dC_t  \\ 
 \end{array}
  \label{eq:dPIHedge}
\end{equation}

\begin{equation}
\begin{array}{l}
 d\Pi _t  + d\Pi _t^{Hedge}  - r^d_t \left( {\Pi _t  + \Pi _t^{Hedge} } \right)dt  \\ 
 \begin{array}{*{20}c}
   {} & {}  \\
\end{array} = \left( {\mathcal{L}^{mkt}\Pi _t  - r^d_t \Pi _t } \right)dt + \beta _t \left( {\mathcal{L}^{mkt}C_t  - r^d_t C_t} \right)dt +  \\ 
 \begin{array}{*{20}c}
   {} & {}  \\
\end{array} \;\;\;\; \left( {\Delta _t^\Pi   + \alpha _t B_t^f + \beta _t \Delta _t^C } \right)S_t \sqrt {v_t } dW_t  + \left( {\vartheta _t^\Pi   + \beta _t \vartheta _t^C } \right)\eta \sqrt {v_t } dV_t  \\ 
 \end{array}
  \label{eq:dPI_dPIHedge}
\end{equation}

Equation (\ref{eq:dPIHedge}) shows the dynamics of the hedging portfolio, where $dB_t^f$ is given by equation (\ref{eq:Bt_f}) and it is assumed that $dB_t = r^d_t B_t dt$. Replacing equations (\ref{eq:dCt}), (\ref{eq:dPI}), (\ref{eq:dPIHedge}) and (\ref{eq:PIHedgeBf}) in equation (\ref{eq:dDiscPI}) and moving the right hand side of equation (\ref{eq:dDiscPI}) to the left, yields equation (\ref{eq:dPI_dPIHedge}). This equation represents the change of value of the total portfolio $\Pi_t^{Tot}$ through time, taking out the domestic risk free return on the total portfolio. This change should be ideally equal to zero according to equation (\ref{eq:dDiscPI}). The right hand side shows two stochastic terms multiplying $dW_t$ and $dV_t$. The hedge ratios $\alpha_t$ and $\beta_t$ should then be chosen to cancel these two terms to remove the uncertainty or randomness of the process. See that the hedge ratios which cancel these two terms are equal to those obtained in section \ref{sec:ImplementationHedging} and given by equation (\ref{eq:AlphaBeta}). The expression $\mathcal{L}^{mkt}f - r^d_t f$ is Heston's partial differential equation (see \cite{Heston1993}) which should be equal to zero for a derivative following the dynamics of the market. See that for the case $f=C_t$, the equation is satisfied and $\mathcal{L}^{mkt}C_t - r^d_t C_t = 0$ because the vanilla option used for hedging is valued according to the assumed market.

\begin{equation}
d\Pi _t^{Tot}  - r_t^d \Pi _t^{Tot} dt = \left( {\mathcal{L}^{mkt}\Pi _t  - r^d_t \Pi _t } \right)dt
  \label{eq:dPItot}
\end{equation}

Replacing $\Pi_t^{Tot} = \Pi_t + \Pi_t^{Hedge}$ in equation (\ref{eq:dPI_dPIHedge}) and eliminating the terms equal to zero yields equation (\ref{eq:dPItot}). If the premium $\Pi_t$ were calculated with the same model assumed for the market, the right hand side of equation (\ref{eq:dPItot}) would be equal to zero and the change of the total portfolio, $\Pi_t^{Tot}$,  minus the interest rate earnings would be zero (the desired result). This means that when the pricing model is not the same model assumed for the market dynamics, the right hand side of equation (\ref{eq:dPItot}) will be different from zero and the evolution of $\Pi_t^{Tot}$ will have a drift. For negative drifts, $\Pi_t^{Tot}$ will show a consistent loss through time. This is what happens in the example of section \ref{sec:DNT}. Section \ref{sec:Faders} shows the opposite case (the drift is positive) where a consistent gain is achieved through time.

\begin{equation}
\Pi _{t_N }^{Tot}  = \Pi _{t_N }  + \Pi _{t_N }^{Hedge} 
  \label{eq:dPItot_tN}
\end{equation}

Consider now equation (\ref{eq:dPItot_tN}), where $\Pi_{t_N}$ represents the final position of a portfolio of contracts (and not cash) and $\Pi_{t_N}^{Hedge}$ is the hedging portfolio whose initial value is equal to $\Pi_{t_0}$ valued with the model under test. Taking expectations, ${\bf E}_{t_0 }^{mkt}[\cdot]$, conditioned on the information at $t_0$ with respect to the risk free measure of the process assumed for the market and considering $P^d_{t,t_N}$ as numeraire (at $t_N$ it is equal to 1) yiels equation (\ref{eq:ExpPi_tN}). This expression gives the expected value of $\Pi_{t_N}^{Tot}$ or what would be lost or earned on average when the model under test is different from the market model. The first term is directly the price of the portfolio $\Pi_{t_N}$ according to the market model (the true value).

\begin{equation}
\begin{array}{l}
 {\bf E}_{t_0 }^{mkt} \left[ {\Pi _{t_N }^{Tot} } \right] = P_{t_0 ,t_N }^d {\bf E}_{t_0 }^{mkt} \left[ {\Pi _{t_N } } \right] + P_{t_0 ,t_N }^d {\bf E}_{t_0 }^{mkt} \left[ {\Pi _{t_N }^{Hedge} } \right] \\ 
 \begin{array}{*{20}c}
   {} & {} & {} & {} & {} & {} \\
\end{array} = \Pi _{t_0 }^{mkt}  - P_{t_0 ,t_N }^d \frac{{\Pi _{t_0 } }}{{P_{t_0 ,t_N }^d }} = \Pi _{t_0 }^{mkt}  - \Pi _{t_0 }  \\ 
 \end{array}
  \label{eq:ExpPi_tN}
\end{equation}

Equation (\ref{eq:ExpHedge}) details the calculation of the expectation of the hedging strategy. This expression is obtained applying the tower law (or iterated expectations) to equation (\ref{eq:PIHedge_tN}) (see that the cash amount ${\rm H}_{t_N}^{t_N}$ cancels the cash value of the position $\alpha_{t_N}$ on underlying and $\beta_{t_N}$ on vanilla call). The expectation on the right hand side is equal to zero, because ${\rm H}_{t_i}^{t_{i-1}}$ over the numeraire $P^d_{t_i,t_N}$ is a martingale as indicated by equation (\ref{eq:ExpH_Pd}), because ${\rm H}_{t_i}^{t_{i-1}}$ is composed of two traded assets: the underlying $S_t B^f_t$ and the vanilla call option $C_t$. Equation (\ref{eq:ExpS_Pd}) shows the martingale condition for the underlying and the vanilla call.

\begin{equation}
{\bf E}_{t_0 }^{mkt} \left[ {\Pi _{t_N }^{Hedge} } \right] = \frac{{ - \Pi _{t_0 } }}{{P_{t_0 ,t_N }^d }} + \overbrace {{\bf E}_{t_0 }^{mkt} \left[ {\sum\limits_{i = 1}^N {\left( {{\bf E}_{t_{i - 1} }^{mkt} \left[ {\frac{{{\rm H}_{t_i }^{t_{i - 1} } }}{{P_{t_i ,t_N }^d }}} \right] - \frac{{{\rm H}_{t_{i - 1} }^{t_{i - 1} } }}{{P_{t_{i - 1} ,t_N }^d }}} \right)} } \right]}^{{\rm Equal \; to \; }0}
  \label{eq:ExpHedge}
\end{equation}

\begin{equation}
{\bf E}_{t_{i - 1} }^{mkt} \left[ {\frac{{{\rm H}_{t_i }^{t_{i - 1} } }}{{P_{t_i ,t_N }^d }}} \right] = {\bf E}_{t_{i - 1} }^{mkt} \left[ {\left( {\alpha _{t_{i - 1} } \frac{S_{t_i } B^f_{t_i}}{{P_{t_i ,t_N }^d }} + \beta _{t_{i-1} } \frac{C_{t_i }}{{P_{t_i ,t_N }^d }} } \right)} \right] = \frac{{{\rm H}_{t_{i - 1} }^{t_{i - 1} } }}{{P_{t_{i - 1} ,t_N }^d }}
  \label{eq:ExpH_Pd}
\end{equation}

\begin{equation}
\begin{array}{*{20}c}
   {{\bf E}_{t_{i - 1} }^{mkt} \left[ {\frac{{S_{t_i } } B^f_{t_i}}{{P_{t_i ,t_N }^d }}} \right] = \frac{S_{t_{i - 1}} B^f_{t_{i-1}}}{{P_{t_{i - 1} ,t_N }^d }}} & {} & {{\bf E}_{t_{i - 1} }^{mkt} \left[ {\frac{{C_{t_i } }}{{P_{t_i ,t_N }^d }}} \right] = \frac{{C_{t_{i - 1} } }}{{P_{t_{i - 1} ,t_N }^d }}}  \\
\end{array}
  \label{eq:ExpS_Pd}
\end{equation}

The conclusion of equation (\ref{eq:ExpPi_tN}) cannot but be overstressed: the average profit or loss of the hedging strategy is the difference of the market price minus the price of the model under test. If the price of the contract was bought at a lower price than the market, a profit should be realized on average. On the other hand, if the position was sold for a lower price than the market, some loss must be realized on average (when a short position is considered, $\Pi_{t_0}$ and $\Pi_{t_0}^{mkt}$ are negative). This conclusion will be verified in section \ref{sec:DNT} for the first case study. See that this conclusion does not depend on the hedging ratios $\alpha_t$ and $\beta_t$ (they could be even random) because the hedging strategy is self-financing (the hedging ratios are fixed along each hedging period). This means that the impact of the hedging ratios are not on the average profit or loss of the hedging strategy, but its uncertainty (e.g. the standard deviation) as will be seen in the coming sections. Therefore, hedge ratios given by a good model close to the market will reduce the uncertainty of the profit or loss and hedge ratios given by another worse model will lead to more dispersion in the distribution of the hedging profit or loss.

When a deal is closed, the average hedging loss can be easily estimated through fair value adjustment (FVA) by comparing the price of the model used to manage the position (the model which gives the hedging ratios) with another better pricing model (perhaps too slow for management) which reflects better the hypothesis of the market. This provision needs to be calculated and accounted for only once when the operation is closed. However, the standard deviation of the hedging loss is considerably more difficult to estimate and may have a bigger impact, depending on the quality of the hedging ratios given by the model. The following sections will estimate this standard deviation of the hedging error.

\section{Definition of relative model risk measure}
\label{sec:ModelRiskMeasure}

The relative model risk of a product priced according to a given model with respect to another reference model is defined by the expected shortfall at a given significance level of the profit and loss given by the hedging strategy of the product priced with its model under the assumption that the market is driven by the dynamics of the reference model. Equation (\ref{eq:MR_ES}) presents the definition of expected shortfall in a formal way, where $X_{p_i ,m_j }^{m_{ref} }$ is the profit and loss of the hedging strategy of product $p_i$, priced with model $m_j$ with respect to the market reference model $m_{ref}$, $\alpha$ is the significance level for which the expected shortfall is calculated and $x_{\alpha}$ is the quantile function given by equation (\ref{eq:x_alpha}).

\begin{equation}
\rho (X_{p_i ,m_j }^{m_{ref} } ) = \frac{{ - 1}}{\alpha }{\bf E}{\kern 1pt} \left[ {X_{p_i ,m_j }^{m_{ref} } {\bf 1}_{\left\{ {X_{p_i ,m_j }^{m_{ref} }  \le x_\alpha  } \right\}} } \right] - \frac{{x_\alpha  }}{\alpha }\left( {\alpha  - P\left[ {X_{p_i ,m_j }^{m_{ref} }  \le x_\alpha  } \right]} \right)
  \label{eq:MR_ES}
\end{equation}

\begin{equation}
x_\alpha   = \inf \left\{ {\left. {x \in R} \right|P\left( {X_{p_i ,m_j }^{m_{ref} }  < x} \right) \ge \alpha } \right\}
  \label{eq:x_alpha}
\end{equation}

In simpler words, the expected shortfall for $\alpha=0.35$ is the expectation of the hedging profit and loss of the worst 35\% paths. See that the expected shortfall is defined as a positive number when losses (negative values of $X_{p_i ,m_j }^{m_{ref} }$) occur.

\begin{equation}
X_{p_i ,m_1 }^{m_{ref} }  \le X_{p_i ,m_2 }^{m_{ref} }  \Rightarrow \rho \left( {X_{p_i ,m_1 }^{m_{ref} } } \right) \ge \rho \left( {X_{p_i ,m_2 }^{m_{ref} } } \right)
  \label{eq:Prop1}
\end{equation}

\begin{equation}
\rho \left( {X_{p_i ,m_1 }^{m_{ref} }  + X_{p_i ,m_2 }^{m_{ref} } } \right) \le \rho \left( {X_{p_i ,m_1 }^{m_{ref} } } \right) + \rho \left( {X_{p_i ,m_2 }^{m_{ref} } } \right)
  \label{eq:Prop2}
\end{equation}

\begin{equation}
\beta  > 0 \Rightarrow \rho \left( {\beta X_{p_i ,m_j }^{m_{ref} } } \right) = \beta \rho \left( {X_{p_i ,m_j }^{m_{ref} } } \right)
  \label{eq:Prop3}
\end{equation}

\begin{equation}
\rho \left( {X_{p_i ,m_j }^{m_{ref} }  + a} \right) = \rho \left( {X_{p_i ,m_j }^{m_{ref} } } \right) - a
  \label{eq:Prop4}
\end{equation}

The main advantage of this measure is the fact that it is coherent (see \cite{Artzner1999}). This means that some properties are satisfied which allow dealing with model risk in a more systematic way. These properties are monotonicity given by equation (\ref{eq:Prop1}), sub-additivity (this property is not satisfied by the measure Value-at-risk and considers that diversification can only reduce risk) by (\ref{eq:Prop2}), positive homogeneity by (\ref{eq:Prop3}) and translation invariance by (\ref{eq:Prop4}).

\section{Case Study I: double-no-touch option}
\label{sec:DNT}

This section applies the hedging methodology of section \ref{sec:Formulation} to estimate the model risk of a double-no-touch (``DNT'') foreign exchange option when the initial valuation and the management throughout the life of the option is carried out using two models with different valuation hypothesis: the volga-vanna (``VV'') and Black Scholes\footnote{This model refers to the valuation using Black Scholes with constant interest rates and a constant volatility equal to the at-the-money volatility chosen at the expiry of the option.} (``BS'') models. A number of conclusions will be reached concerning model risk (what happens when a limited model is used to quote and manage a deal) and provision calculation. The selected framework is very appropriate to compare different models because the market dynamics are known (Heston's model) and therefore the correct price of the product. The comparison among models is homogeneous and provides an objective criterion to judge and measure which model is better.

\begin{figure}[htbp]
	\centering
	\includegraphics[width=0.450\textwidth]{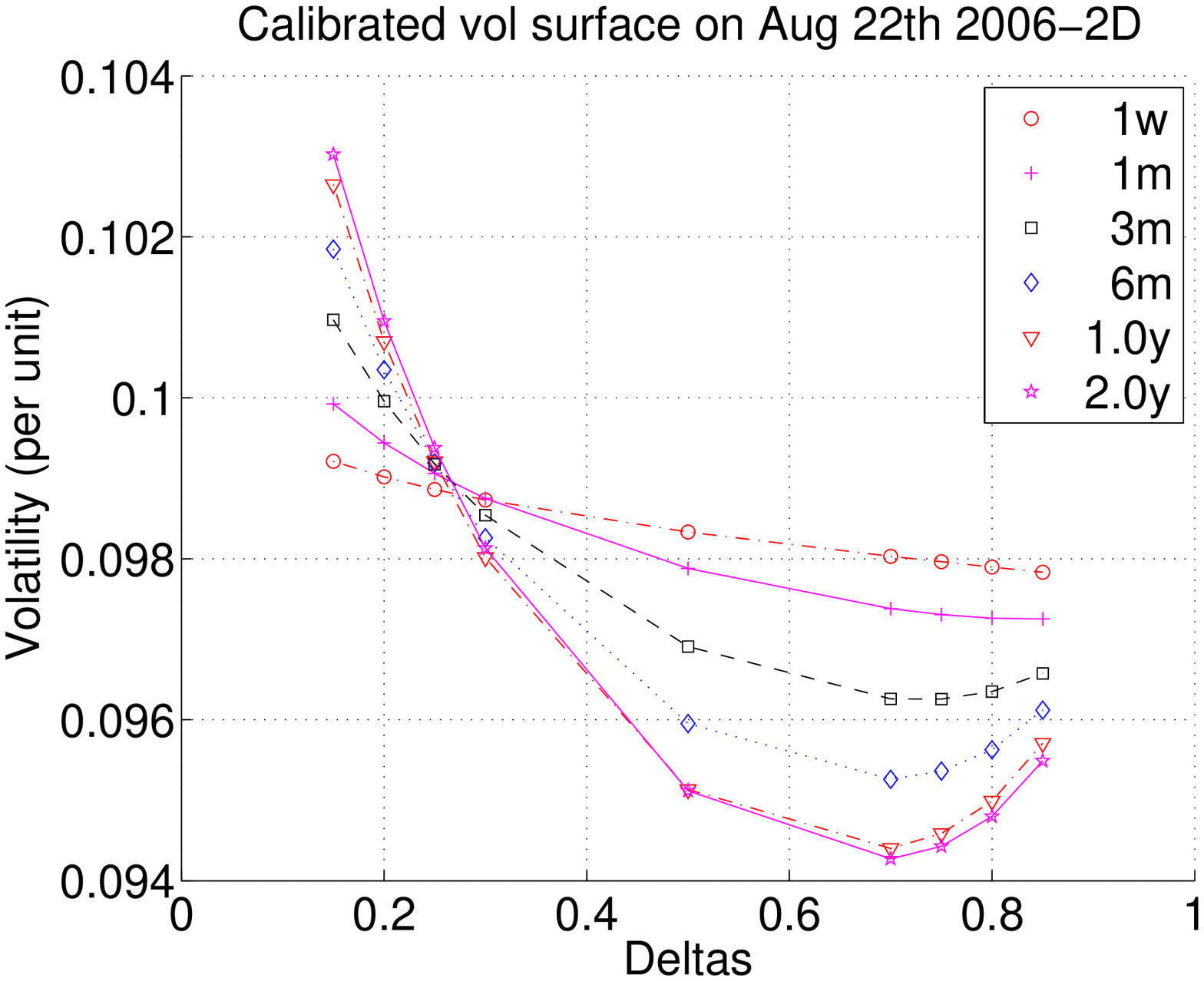}
	\includegraphics[width=0.450\textwidth]{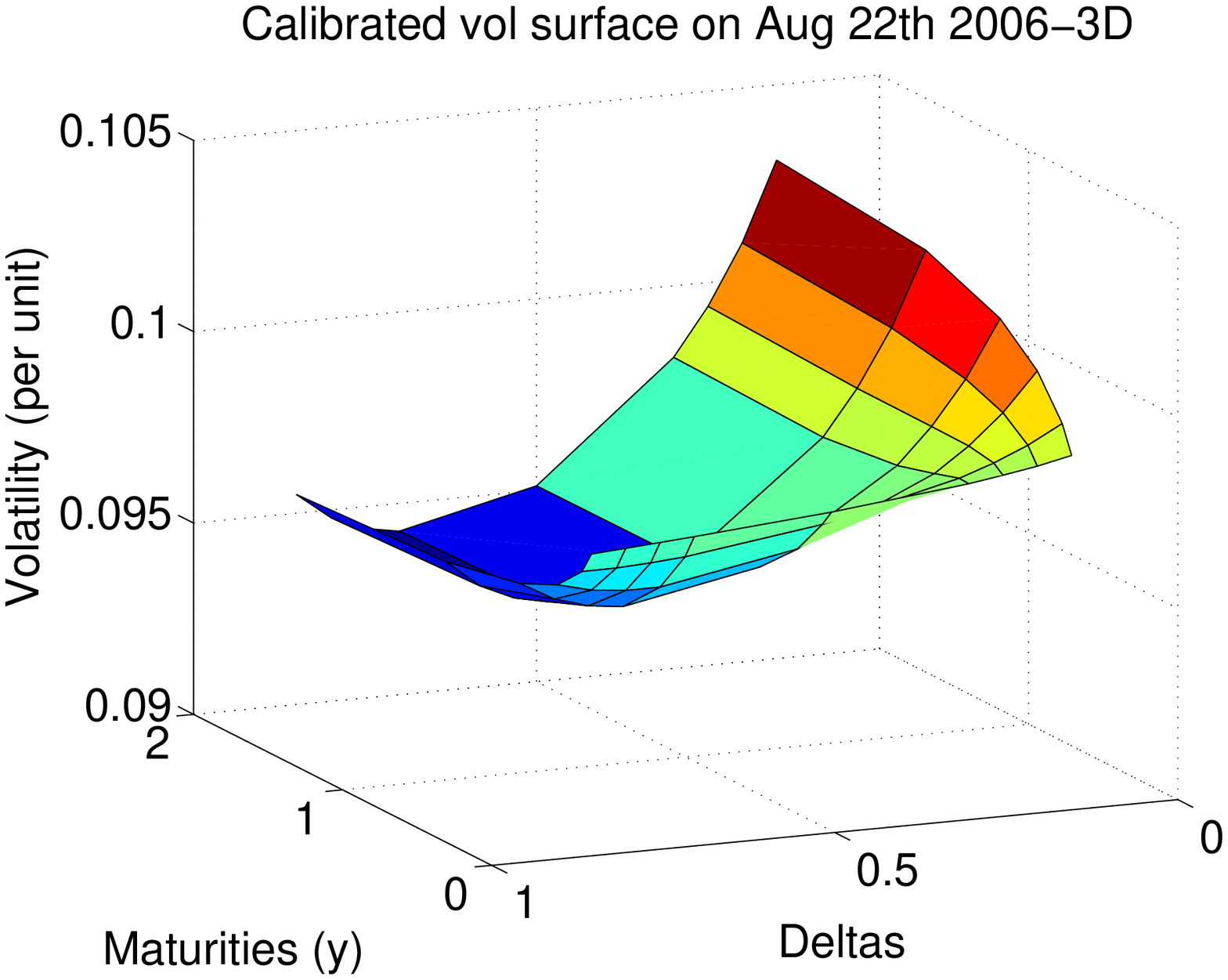}	
	\caption{Heston's volatility surface calibrated to 1 year market on Aug 22nd 2006 with $v_0 = 0.0097$, $\kappa = 1.1$, $\theta = 0.0097$, $\eta = 0.14$ and $\rho = 0.14$.}
	\label{fig:CalibratedVolSurf}
\end{figure}

The deal under test is a double-no-touch option maturing 1 year and 17 days from present time. The underlying is the EUR/USD foreign exchange rate. This contract pays the notional in USD at maturity provided that the underlying does not breach two barrier levels (1.2130 and 1.3622). If any of the barriers get touched, the product cancels and pays out zero. Figure \ref{fig:CalibratedVolSurf} shows the volatility surface of the EUR/USD generated by Heston's model with parameters calibrated to market on August 22nd, 2006 for a 1 year horizon. The market is skewed favoring out-of-the-money calls (they have higher volatility). The spot price is 1.2812 and interest rates have been set to zero (both domestic and foreign).

\begin{figure}[htbp]
	\centering
	\includegraphics[width=0.45\textwidth]{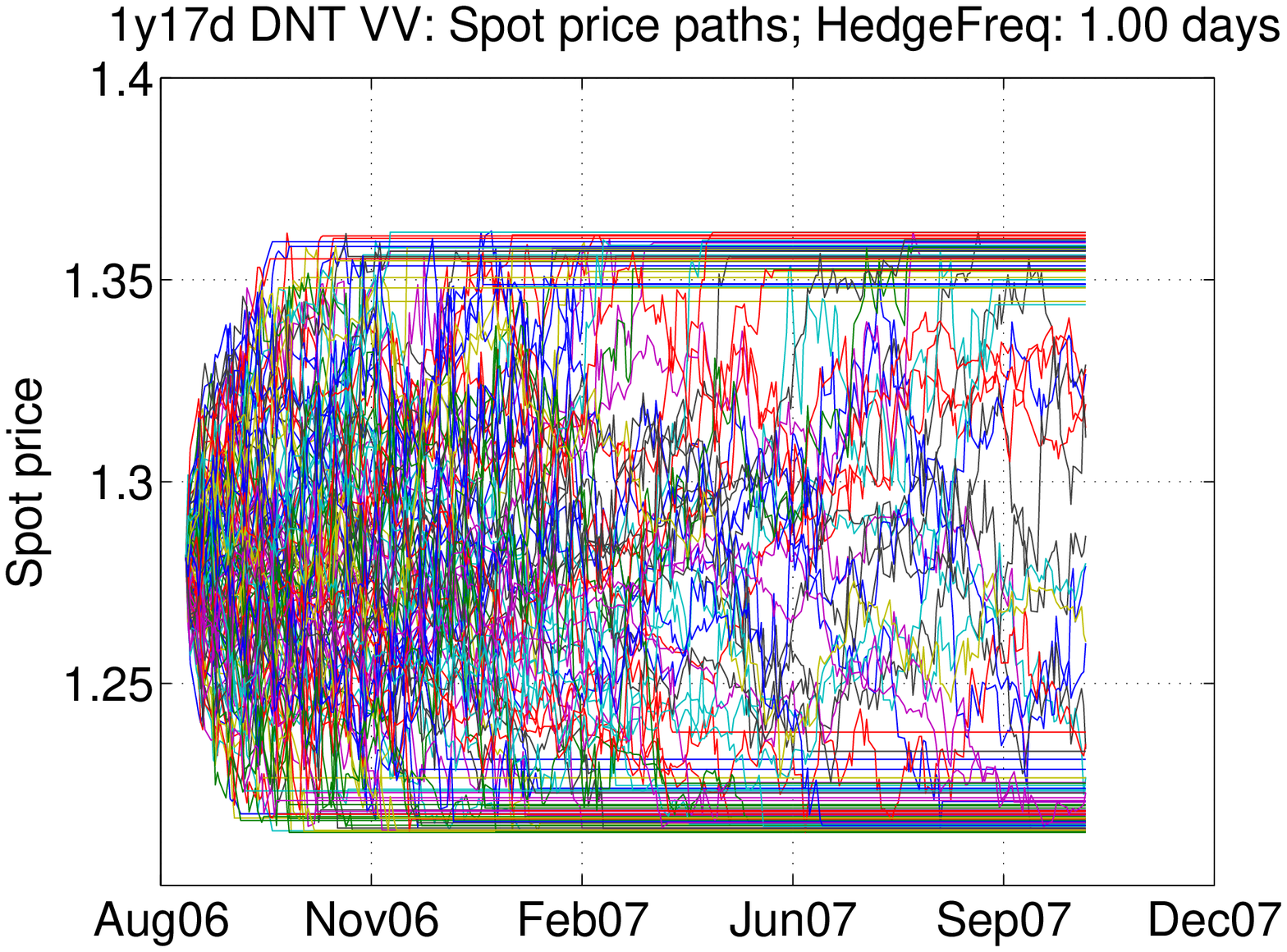}
	\includegraphics[width=0.45\textwidth]{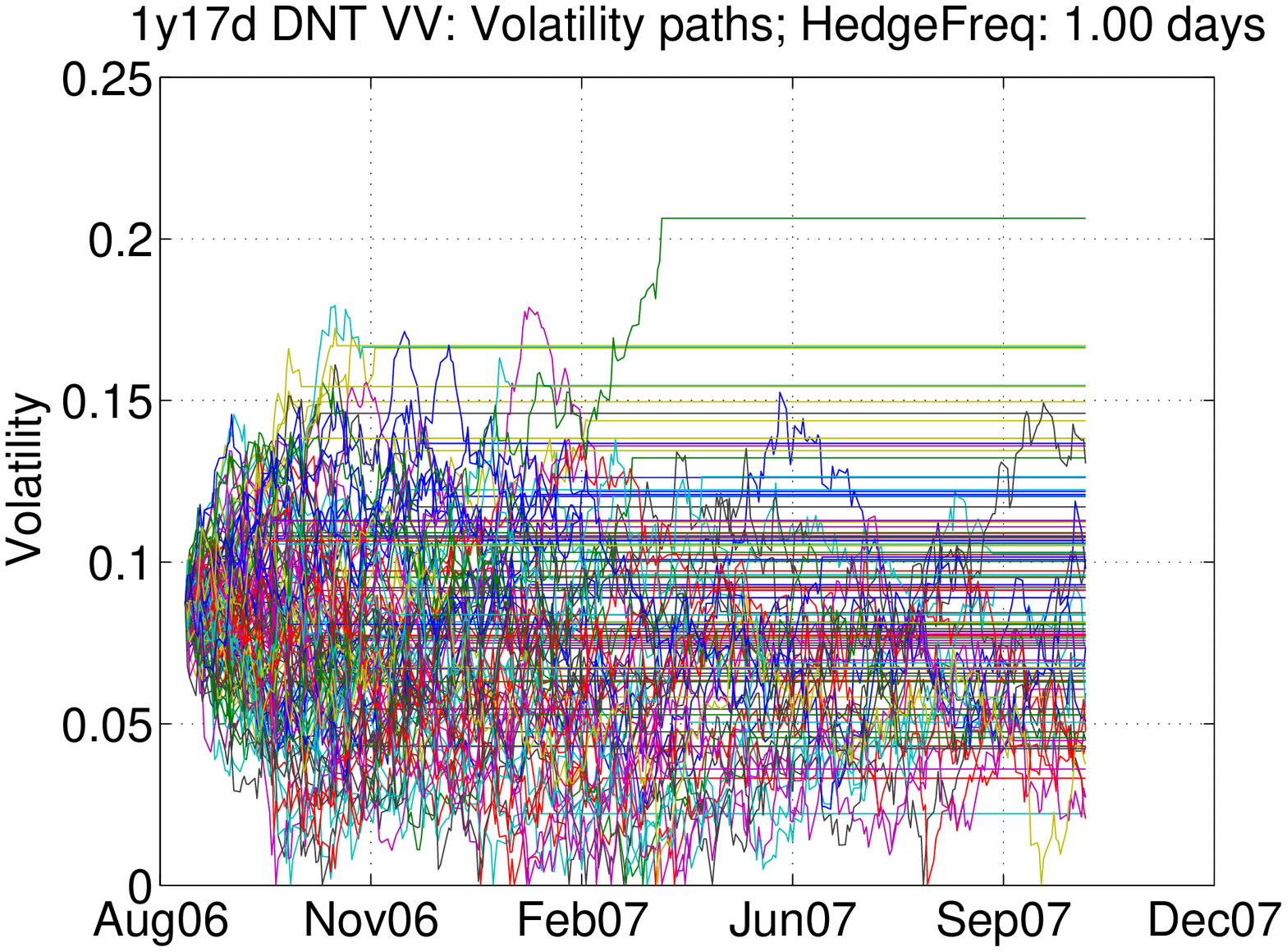}
	\caption{Underlying ($S_t$) and volatility ($\sqrt{v_t}$) paths simulated according to Heston's model. Straight lines show a barrier breach on the following date.}
	\label{fig:DNTspot_vol}
\end{figure}

Figure \ref{fig:DNTspot_vol} shows the evolution of the underlying $S_t$ and volatility ($\sqrt{v_t}$) for 120 simulated paths according to Heston's model with the parameters indicated in figure \ref{fig:CalibratedVolSurf}. When the barriers get touched, the path ends (the deal disappears) and the last live spot or volatility point is maintained up to the end of the contract drawing a straight line. The few straight lines which are not too close to the barrier levels are due to daily big movement events (up to 3 standard deviation).

\begin{table}[htbp]
  \centering
    \begin{tabular}{ccc}
    VV     &    BS   & Heston \\
    \hline
    0.0839 &  0.0466 & 0.1122 \\
    \end{tabular}
  \caption{DNT premium per unit of notional according to different models.}
  \label{tab:PremiumDNT}
\end{table}

Table \ref{tab:PremiumDNT} shows the premium per unit of notional in USD according to the three models under consideration. Heston's price has been calculated by the same Monte Carlo method used to simulate the hedging paths (same Heston's parameters) and applying a brownian bridge to account for the probability of touching the barriers in between two consecutive discrete observations (the volatility used for the brownian bridge in each interval is the average of $\sqrt{v_t}$ at the two time extremes of the interval). A total number 120 thousand paths were simulated. Model risk is significant given how prices differ from each other. Heston's price is about 3 times the Black Scholes price and 33\% higher than volga-vanna price. Assuming that the market is driven by Heston's model, the correct price of the deal is given by Heston's price.

\begin{figure}[htbp]
	\centering
	\includegraphics[width=0.45\textwidth]{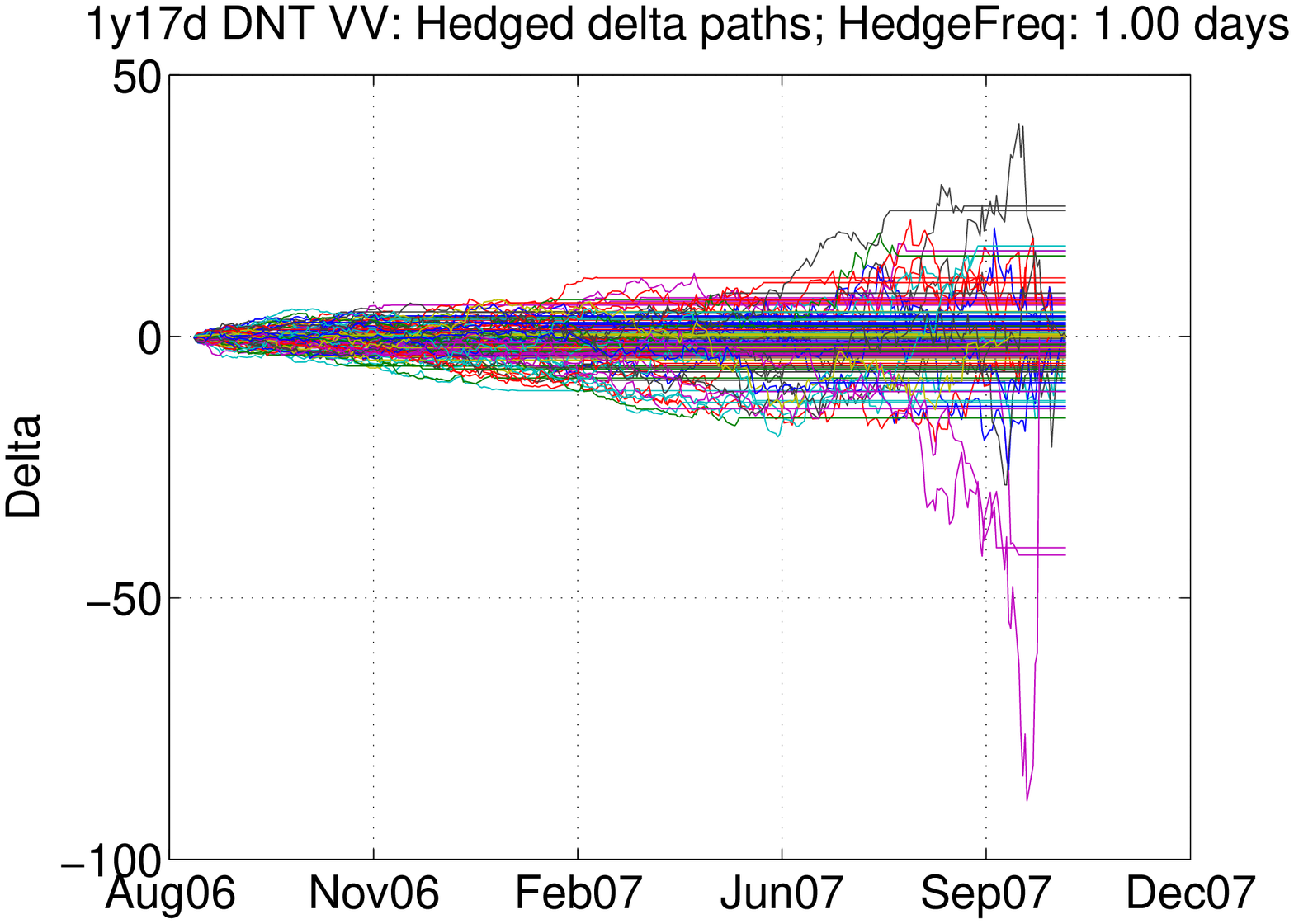}
	\includegraphics[width=0.45\textwidth]{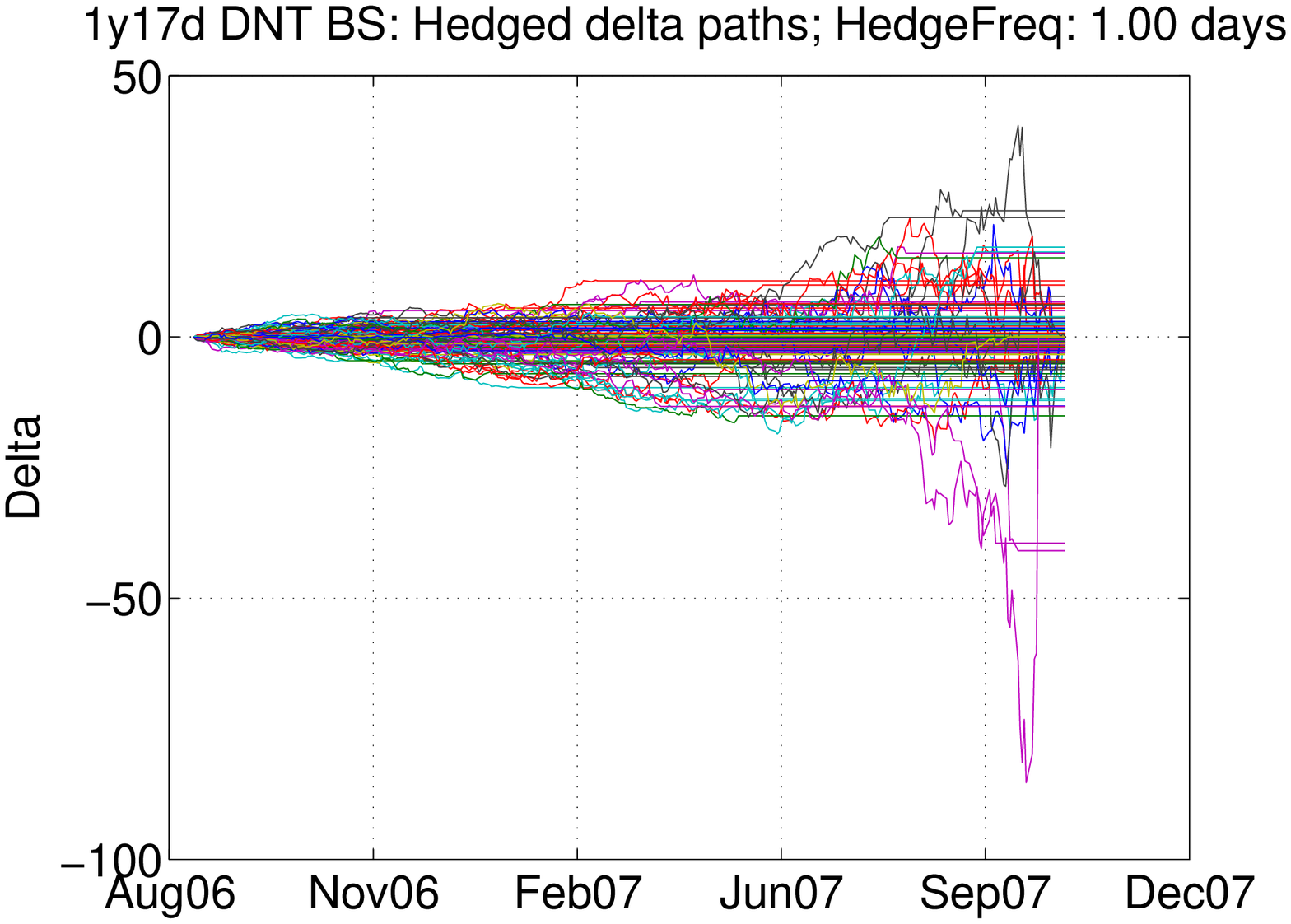}
	\includegraphics[width=0.45\textwidth]{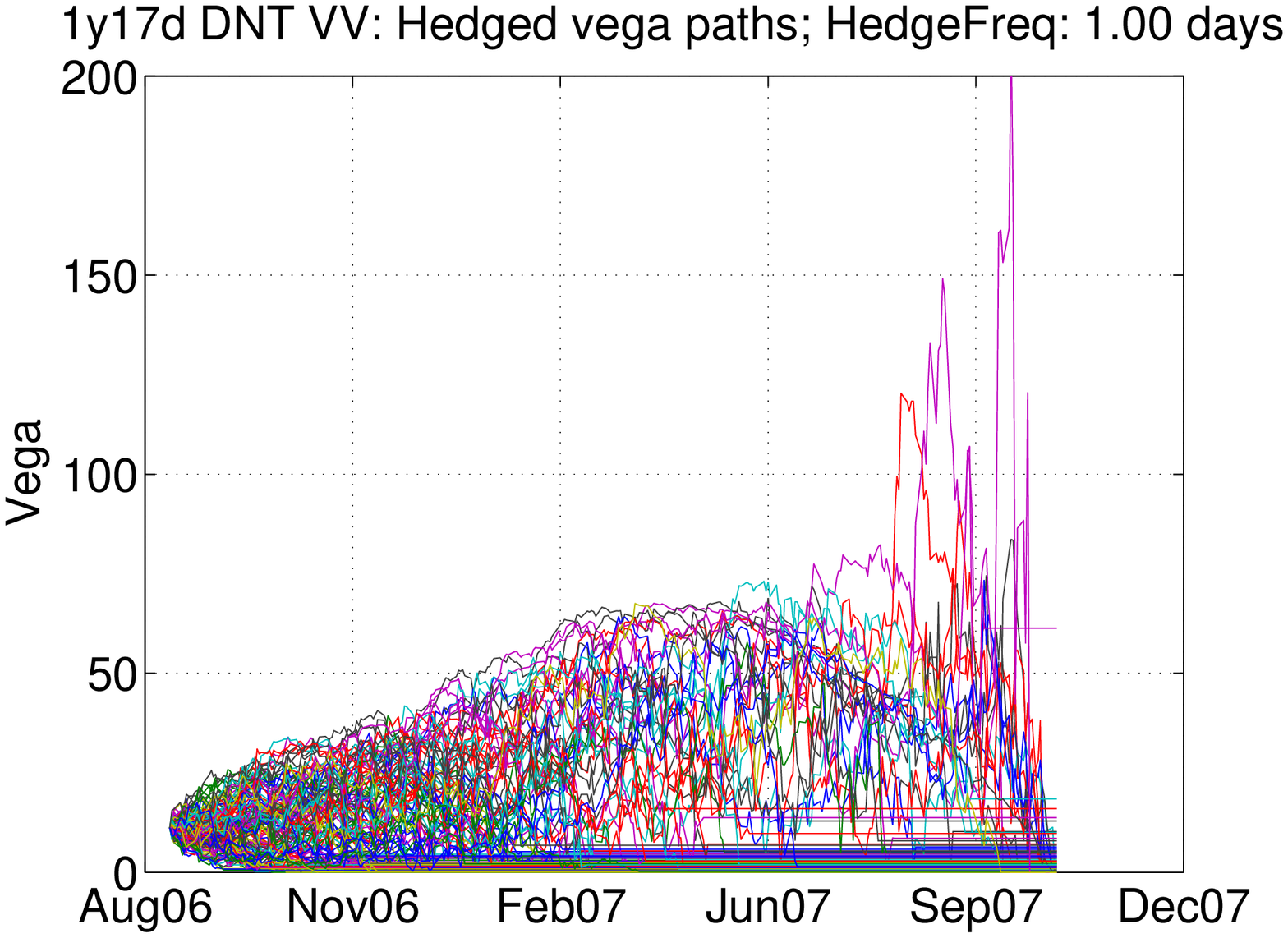}
	\includegraphics[width=0.45\textwidth]{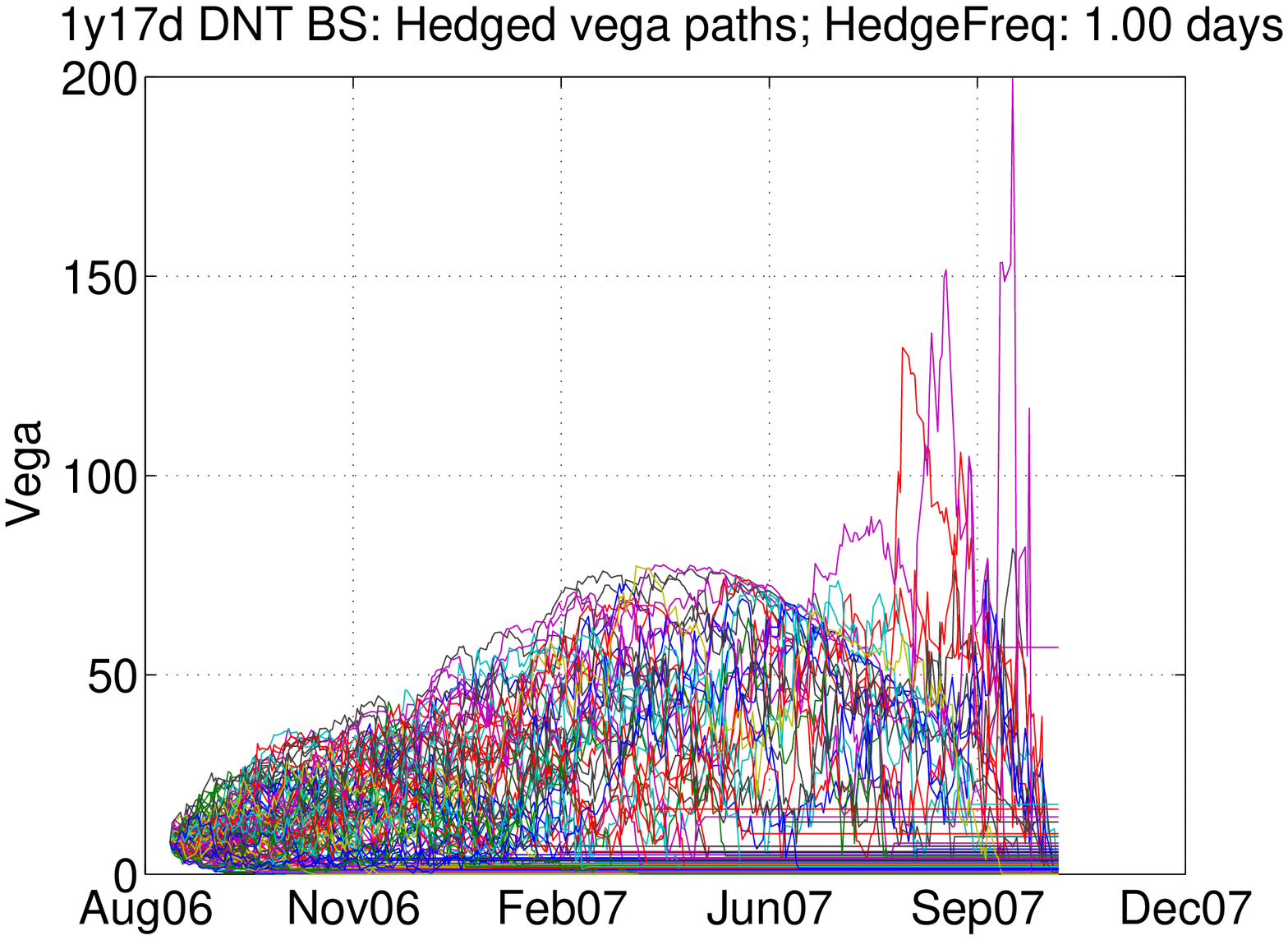}
	\caption{Delta paths (upper plots) and vega paths (lower plots) for volga-vanna (left plots) and Black Scholes (right plots) for a 1y double-no-touch option -with barriers 1.2130 and 1.3622.}
	\label{fig:DNTdelta_vega}
\end{figure}

Figure \ref{fig:DNTdelta_vega} shows the hedge ratios for both the volga-vanna (``VV'') and Black Scholes (``BS'') models. The same set of paths presented in figure \ref{fig:DNTspot_vol} has been used both models to allow for better comparison. The upper plots show the delta for ``VV'' (left) and ``BS'' (right) models. The lower plots show the premium sensitivity with respect to Heston's initial variance $v_t$ (in the plot it is quoted as ``vega''), again for ``VV'' (left) and ``BS'' (right) models. The premium profile of the double-no-touch varying spot is similar to a semi-circle. Near the barriers the premium is equal to zero and in between the barriers the premium reaches its maximum. Therefore, in the middle point delta is close to zero, near the upper barrier delta is negative and near the lower barrier delta is positive. That is why delta paths start near zero and they progressively get either positive or negative depending on whether the spot approaches the lower or upper barrier levels. See that near expiry deltas can be very big when the spot price gets near the barrier levels. This is one of the major hedging problems of barrier options near expiry.

\begin{figure}[htbp]
	\centering
	\includegraphics[width=0.45\textwidth]{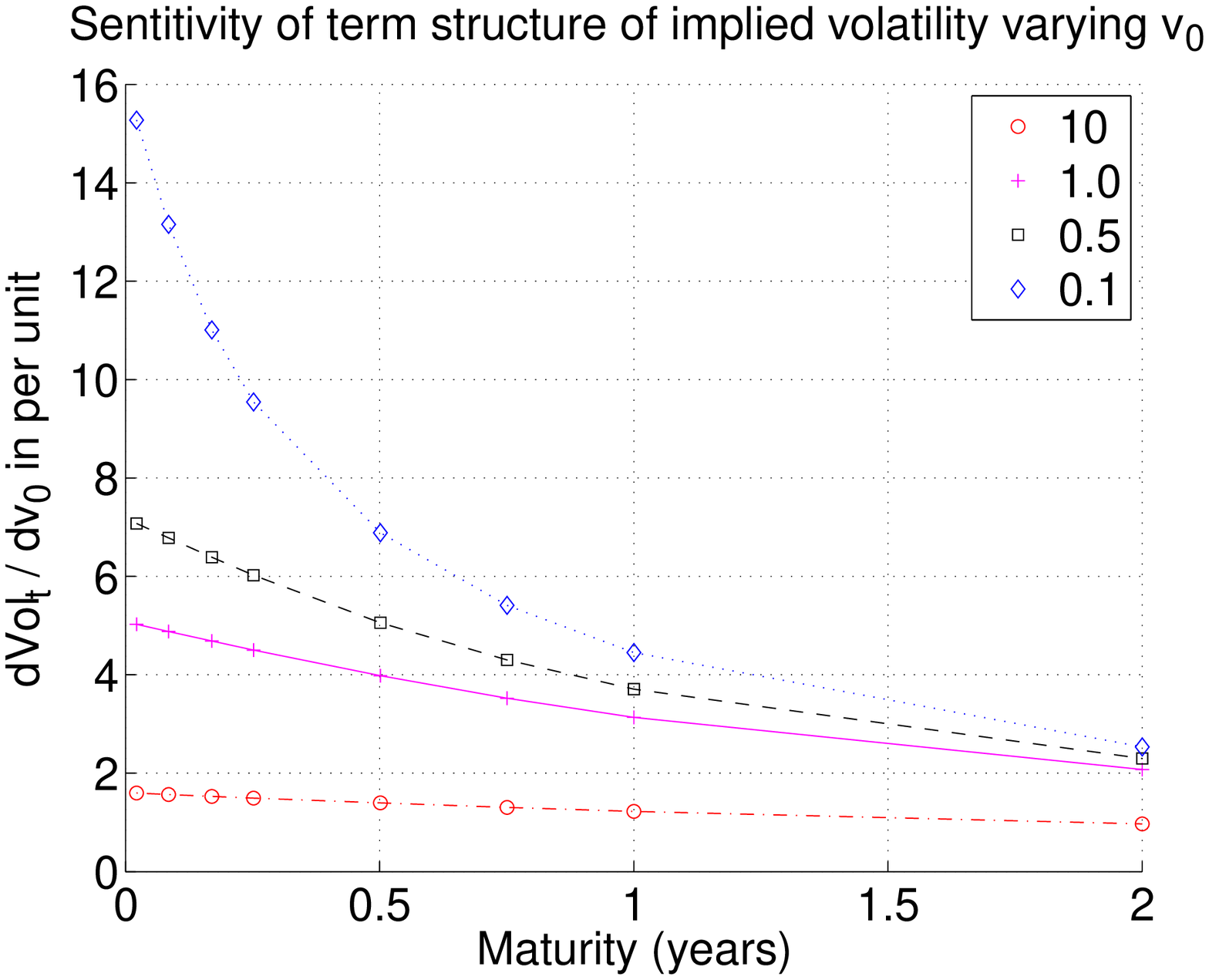}
	\includegraphics[width=0.45\textwidth]{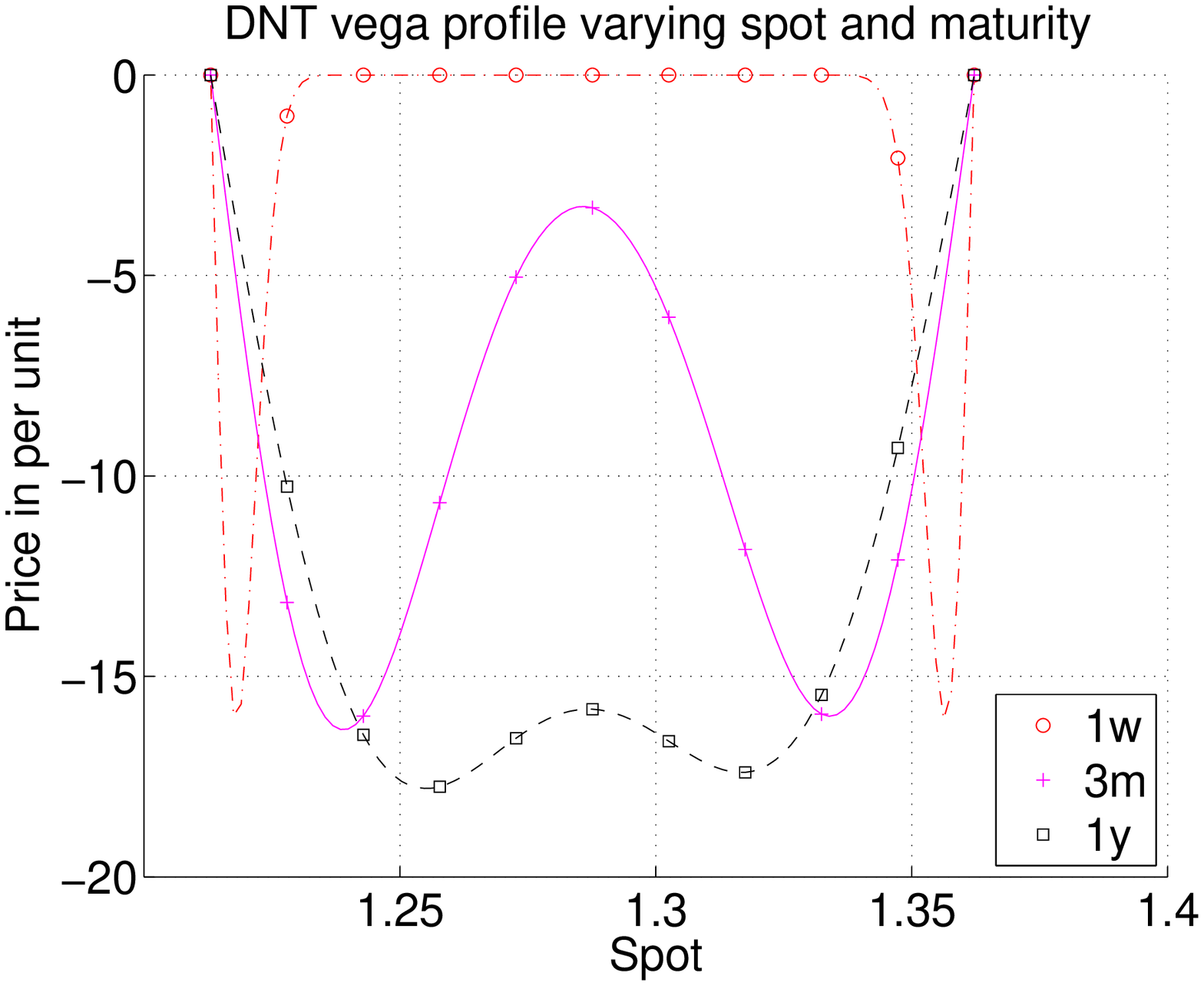}
	\caption{Per unit sensitivity of Heston's implied volatility with respect to initial variance $v_t$ (left) and of Black Scholes DNT vega varying spot and initial variance as a factor (10, 1, 0.5 and 0.1) of long term variance $\theta$ (right).}
	\label{fig:DNTBigVega}
\end{figure}

The lower plots of figure \ref{fig:DNTdelta_vega} show the premium sensitivity with respect to the initial variance, $\vartheta$, computed according to equation (\ref{eq:GreekDefinition}). The sensitivity $\vartheta$ should be negative (an increase in volatility rises the chances to touch the barriers and the price decreases). It appears positive because a short position is held. See that approaching expiry, big values of $\vartheta$ are reached for a few paths. If equation (\ref{eq:vega_model}) is considered to calculate $\vartheta$, the big levels are explained by the fact that near the barriers and close to expiration, the option vega may get rather high and so the sensitivity of Heston's implied volatility to movements of $v_t$ for low values of $v_t$. This is shown in figure \ref{fig:DNTBigVega} created for the path which reaches the 200 vega level. This level is attained 18 days before expiry with a spot level of $S_t = 1.2193$ and a variance level $v_t = 8.1894\cdot 10^{-4}$ (a volatility level of 2.86\%). The left plot shows the sensitivity of the term structure of implied volatility changing the initial variance when the initial variance varies by a factor (10, 1, 0.5 and 0.1) of the long term variance $\theta$ (for the path considered this factor is 0.0844 and the sensitivity is around 15). The right plot shows the Black Scholes double-no-touch vega varying spot and time to expiry for a volatility level of 2.86\% and zero interest rates. This plot presents how vegas concentrate close to the barriers near expiry (for the given path, the vega contribution is around 14). If both contributions are multiplied according to equation (\ref{eq:vega_model}), the observed value is obtained: $\vartheta = 15 \cdot 14 = 210$.

\begin{figure}[htbp]
	\centering
	\includegraphics[width=0.45\textwidth]{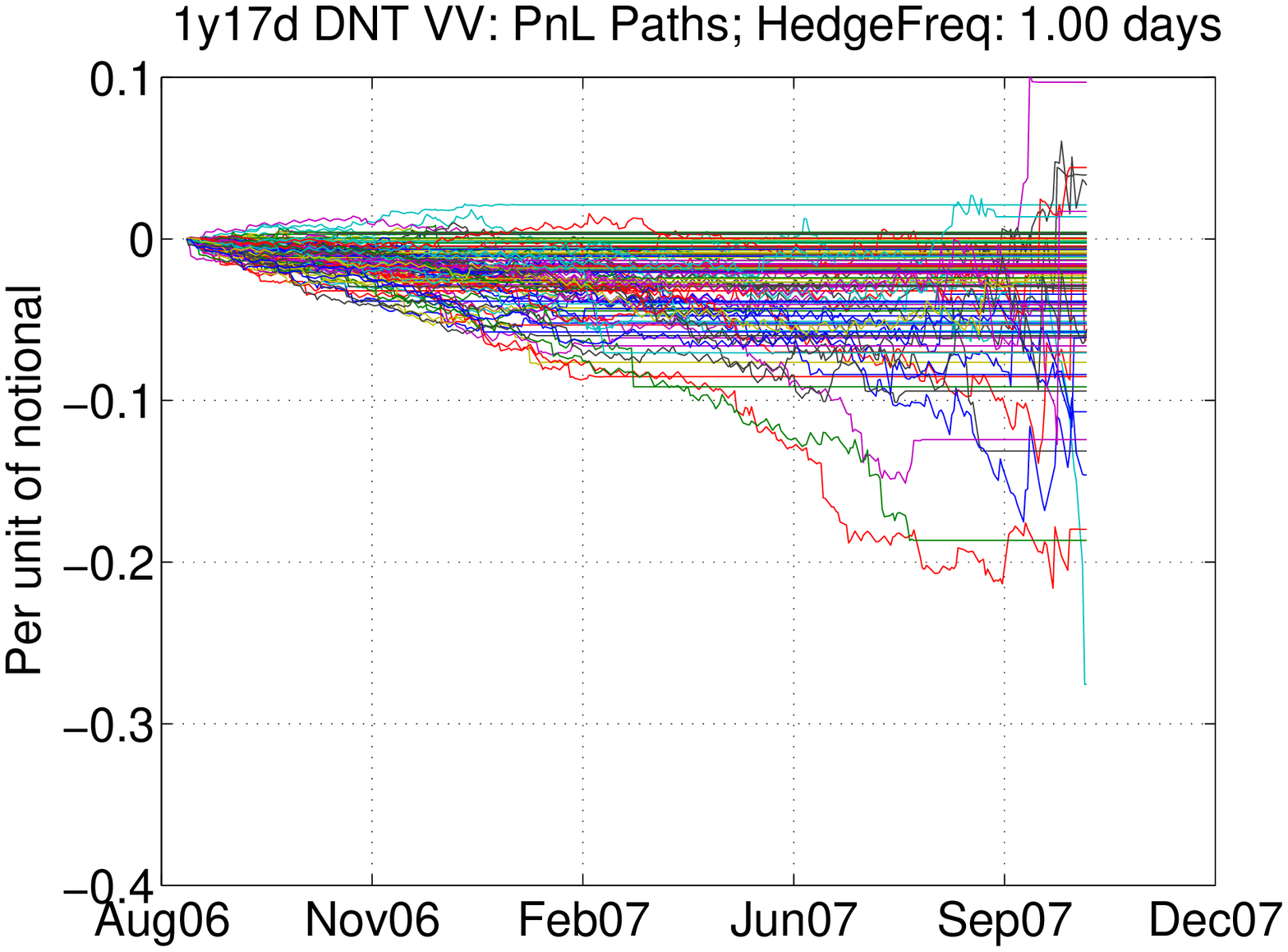}
	\includegraphics[width=0.45\textwidth]{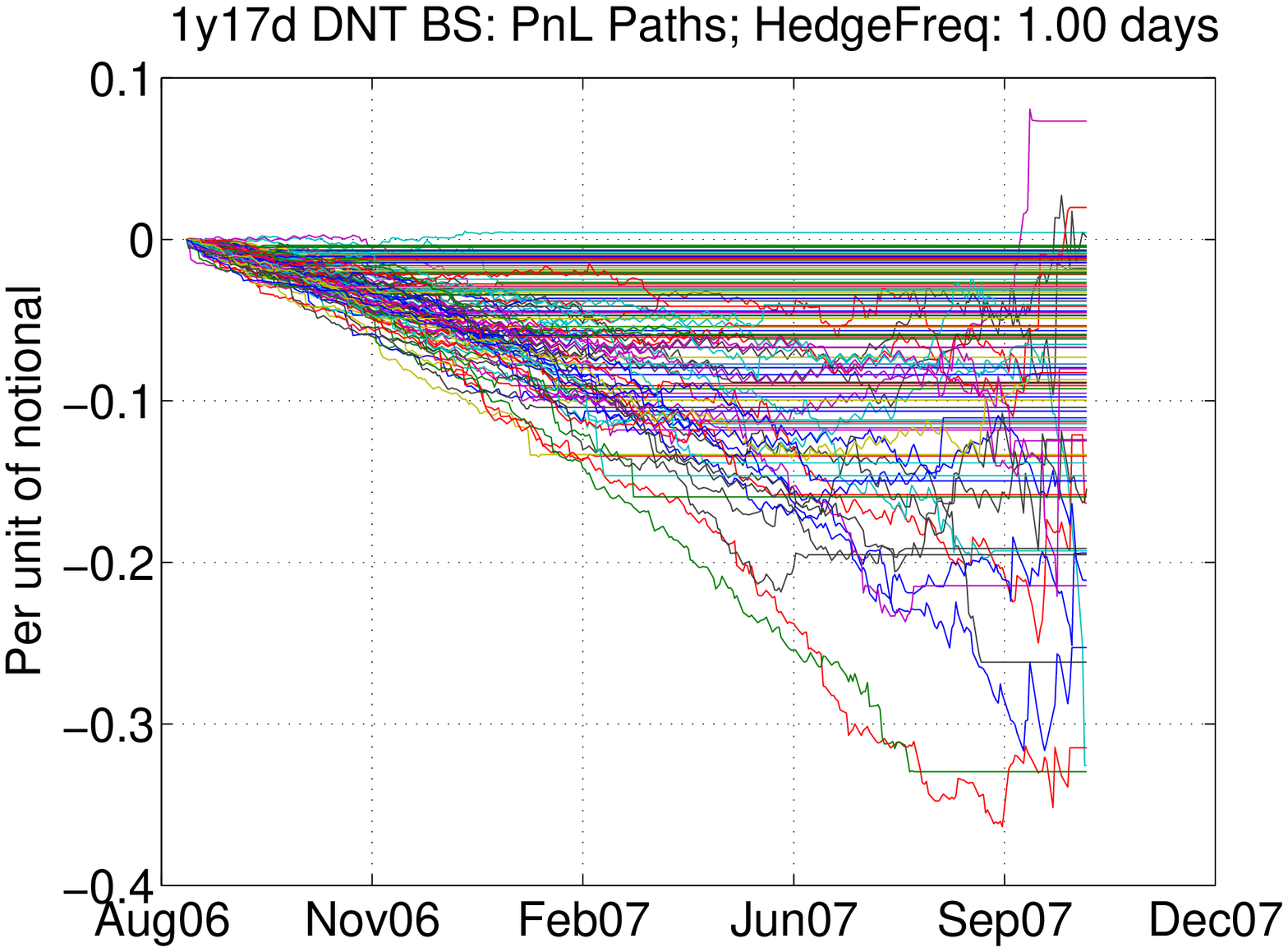}
	\caption{Comparison of P\&L paths hedged with volga-vanna (left) and Black Scholes (right).}
	\label{fig:DNT_PnL_VV_BS}
\end{figure}

\begin{figure}[htbp]
	\centering
	\includegraphics[width=0.45\textwidth]{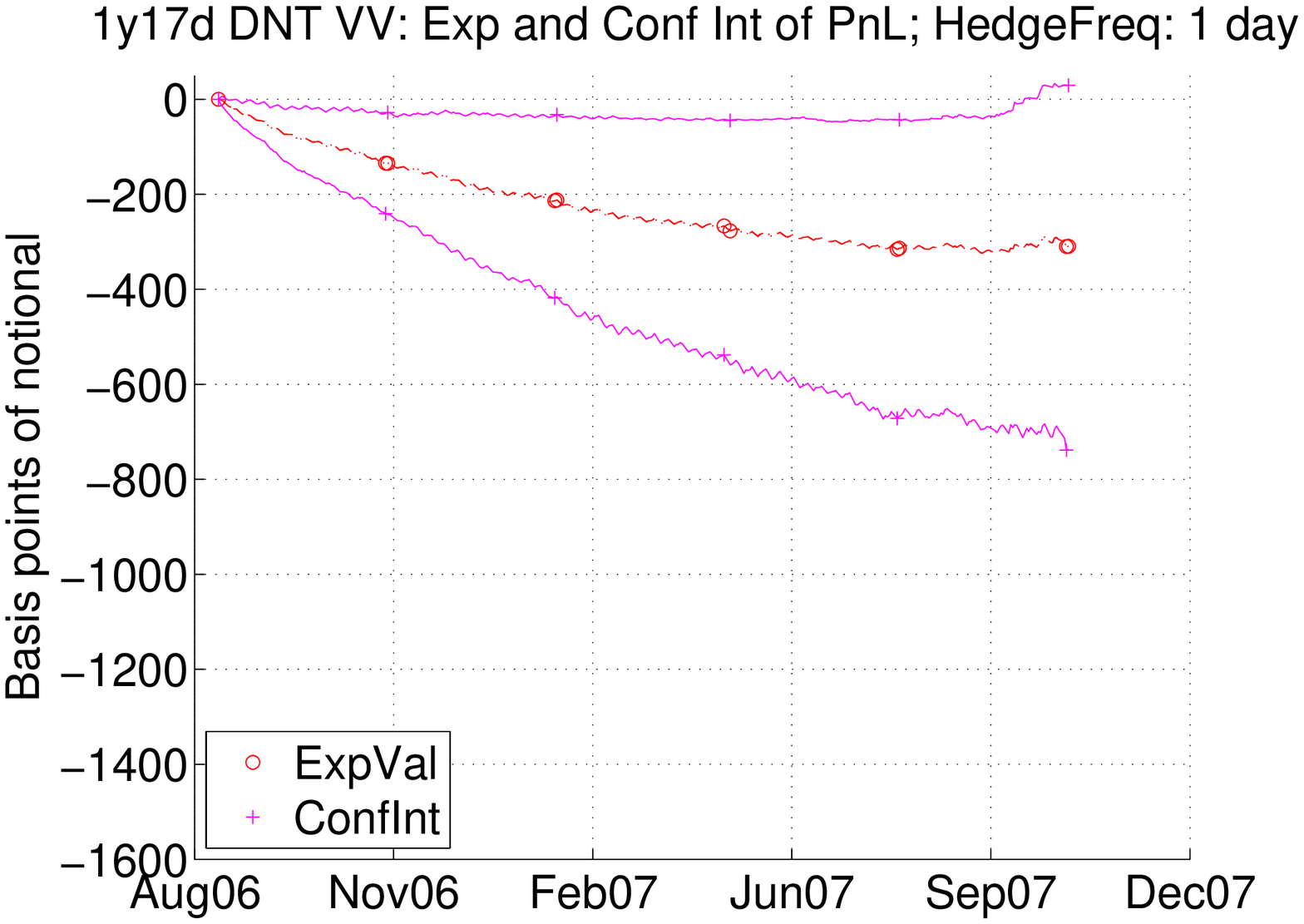}
	\includegraphics[width=0.45\textwidth]{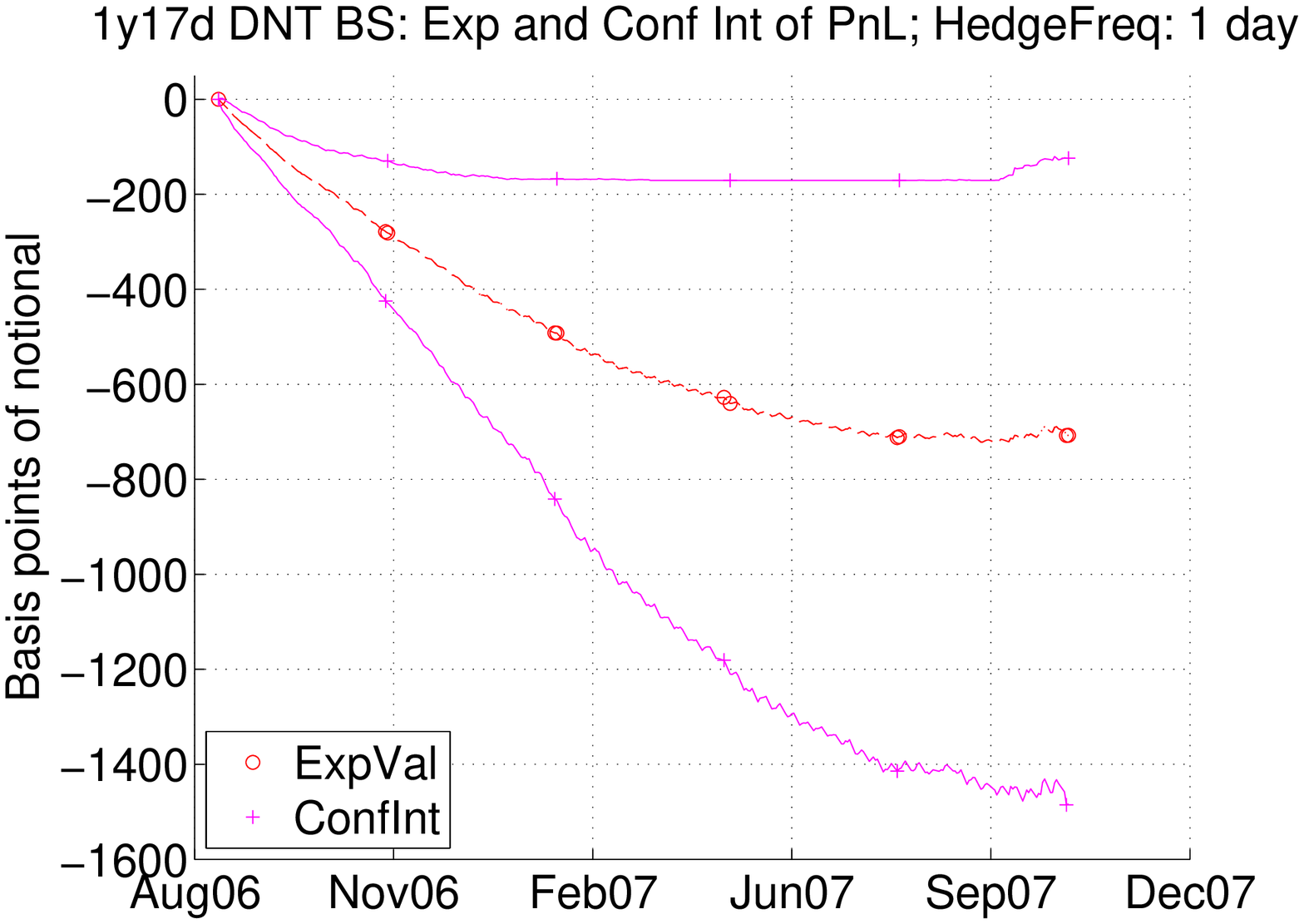}
	\caption{Comparison of confidence intervals at 35\% significance of the P\&L hedged with volga-vanna (``VV'' left) and Black Scholes (``BS'' right) models.}
	\label{fig:DNT_ExpVal_StdDev}
\end{figure}

Figure \ref{fig:DNT_PnL_VV_BS} displays the evolution of the hedging earnings per unit of notional, $\Pi^{Tot}_{t}$, calculated according to equation (\ref{eq:PIHedge_tN}) for the volga-vanna model (left) and the Black Scholes model (right). It is very remarkable that according to figure \ref{fig:DNTdelta_vega} the hedge ratios given by the volga-vanna model (left plots) and the Black Scholes model (right plots) are almost indistinguishable, whereas the evolution of the hedging earnings presented in figure \ref{fig:DNT_PnL_VV_BS} are very different. Both models show consistent hedging losses (negative earnings) throughout the life of the option. Compared to the Black Scholes model, the volga-vanna model shows less losses on average and they are more concentrated (they are less uncertain).

Figure \ref{fig:DNT_ExpVal_StdDev} shows the evolution through time of the expected value (``ExpVal'') and two confidence intervals (``ConfInt'') of the hedging loss for the volga-vanna (left) and Black Scholes (right) models. The confidence intervals are calculated as the expectation of the tails of the distribution with a degree of significance of 35\%. The upper limit calculates through time the expectation of the profit and loss paths above the 65\% percentile (the average of the best 35\% hedging paths) and the lower limit is the expectation of the profit and loss paths below the 35\% percentile (the average of the worst 35\% hedging paths). The lower confidence limit is the negative of the expected shortfall defined in section \ref{sec:ModelRiskMeasure}, for a significance level of 35\%. Depending on how small this degree of significance is selected, the profile of the user would become more risk averse. If both models are compared, it is very clear that the volga-vanna model has lower expected loss with less uncertainty. The uncertainty of the losses can be measured by the difference between the expectation of the hedging loss and the negative of the expected shortfall (lower limit of the confidence interval). Figure \ref{fig:DNT_ExpVal_StdDev} shows that the profit and loss distribution is rather symmetric (the expected hedging loss is rather in between the confidence interval). However, there is no argument to ensure this fact.


\begin{table}[htbp]
  \centering
    \begin{tabular}{ccc}
    VV    &       & BS \\
    0.0839  & Initial model price: $P_0$ & 0.0466 \\
    0.0309 & Expected hedging cost: $EHC$ & 0.0707 \\
    \hline
    0.1148 & Price including hedging cost: $P_0 + EHC$ & 0.1173 \\
    0.1122 & Heston's price & 0.1122 \\
    \hline
    0.0736 & Model Risk = expected shortfall & 0.1483 \\
    \hline
    0.1575 & Final price: $P_0 + MR$ & 0.1949 \\
    \end{tabular}
  \caption{Compared summary of VV and BS in terms of pricing and hedging.}
  \label{tab:SummaryDNT}
\end{table}

Table \ref{tab:SummaryDNT} summarizes and interprets the study carried out in this section. It presents quantities calculated with the volga-vanna (left hand side) and Black Scholes (right hand side) models. All quantities are expressed in USD per unit of notional. The first row shows the initial premium. The second column presents the expected hedging cost taken from the path in between the confidence intervals of figure \ref{fig:DNT_ExpVal_StdDev}. The third row shows the price which should have been charged to the client in order not to loose money on average when closing the deal: the initial premium plus the expected hedging cost ($EHC$). The forth row shows the correct price for the deal (Heston's price). Now, equation (\ref{eq:ExpPi_tN}) can be verified by the fact that the price which should have been charged to the client according to both models and the market price given by Heston's model agree very well (having into account that only 120 paths have been simulated). This obvious fact is expressed mathematically by equation (\ref{eq:Pmkt}) saying that the market price should be equal to the model price plus the expected hedging cost.

\begin{equation}
P_{{\rm mkt}}  = P_{{\rm model}}  + {\rm Hedging \: cost}
  \label{eq:Pmkt}
\end{equation}

Rewriting equation (\ref{eq:Pmkt}) leads to equation (\ref{eq:Lemma}) which states that the expected hedging cost (loss) is given by the premium difference between the model assumed for the market and the model used for hedging. This equation is the same as equation (\ref{eq:ExpPi_tN}) and it justifies that whatever is the unknown model for the market, getting apart from it is worrying because it directly translates into potential hedging losses on average.

\begin{equation}
{\rm Hedging \: cost} = P_{{\rm mkt}}  - P_{{\rm model}}
  \label{eq:Lemma}
\end{equation}

The expected hedging cost gives a reference of what on average might be lost. However, it might be necessary to add a cushion to consider the uncertainty of this loss (the difference between the expected loss and the negative of the expected shortfall). The fifth row of table \ref{tab:SummaryDNT} shows the relative measure of model risk introduced in section \ref{sec:ModelRiskMeasure} (the expected loss plus the cushion) for a 35\% significance level. See that the Black Scholes model has almost twice as much model risk than the volga-vanna model for this particular deal with a 35\% significance level. Finally, the last row of table \ref{tab:SummaryDNT} shows the final price which should have been given to this deal in order to be sufficiently protected (the initial price plus the expected shortfall). See that this final price is twice the price given by the volga-vanna model and more than four times the price given by the Black Scholes model. In this case, it is obvious that hedging with the volga-vanna model would be better than using Black Scholes. However, the volga-vanna model still has a remarkable amount of model risk.

\section{Case Study II: portfolio of faders}
\label{sec:Faders}

This section applies the same methodology to estimate model risk and calculate provisions for a portfolio of fader options when they are quoted and managed with the volga-vanna model. It will be seen that under the same assumptions the particular deals traded have very little model risk. A fader is a call or put option with a contingent notional. A number of monthly or weekly fading dates are considered between present time and the expiry of the option. An upper (``ULEVEL'') and lower (``DLEVEL'') fading level is defined for each fading date (they are usually equal for all of them). The initial notional is equal to zero. When the underlying spot is in between the two fading levels on a given fading date, a fraction of the maximum notional is added to the current notional. If the spot goes in between the fading levels for all fading dates, the resulting product is a vanilla option with the maximum notional. In addition, it is possible to define knock-out barriers (they are continuously monitored). When the barriers are breached, the fader will no longer add any more notional. This helps to protect one of the counterparts in case the option gets highly in-the-money against them.

\begin{figure}[htbp]
	\centering
	\includegraphics[width=0.3\textwidth]{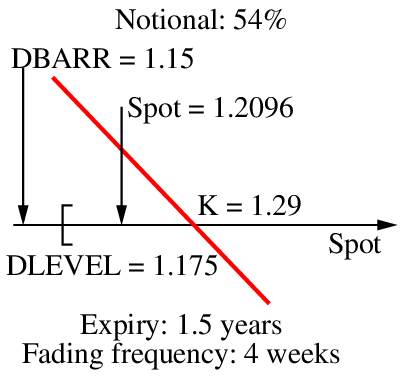}
	\includegraphics[width=0.3\textwidth]{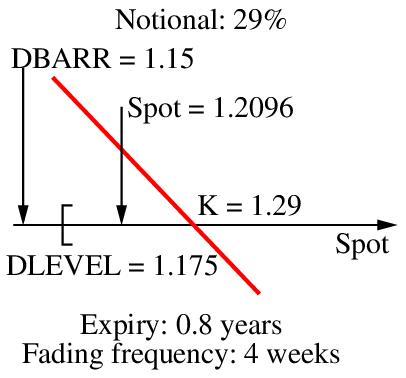}
	\includegraphics[width=0.3\textwidth]{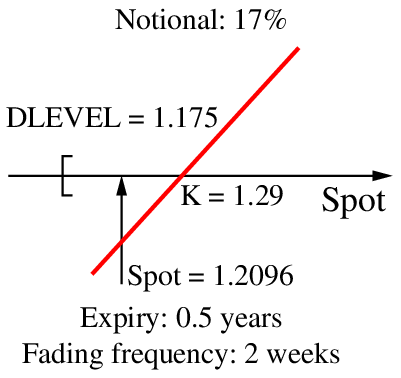}
	\caption{Comparison of the expected value (left) and standard deviation (right) of P\&L for volga-vanna (``VV'') and Black Scholes (``BS'') models.}
	\label{fig:faders}
\end{figure}

Figure \ref{fig:faders} shows the payout functions (bold red lines) varying spot (horizotal axis) of three fader deals which qualitatively represent a portfolio. The underlying is the EUR/USD exchange rate with spot $S_0 = 1.2096$. Each deal is in reality composed by a fader call and put with the same strike to build a forward contract with contingent notional. The left plot shows a short position of a forward fader with strike 1.29, a lower fading level of 1.175 (there is no upper fading level or it is very high), a lower barrier at 1.15 (there is no upper barrier) and monthly fading dates. The notional is 54\% and it expires in 1.5 years. The middle plot of figure \ref{fig:faders} is the same operation but with an expiry of 0.8 years and a notional of 29\%. Finally, the right plot shows a long position on a forward fader without barriers, with strike 1.29, lower fading level of 1.175 and fading dates every two weeks, with a notional of 17\% and expirying in 6 months.

\begin{figure}[htbp]
	\centering
	\includegraphics[width=0.45\textwidth]{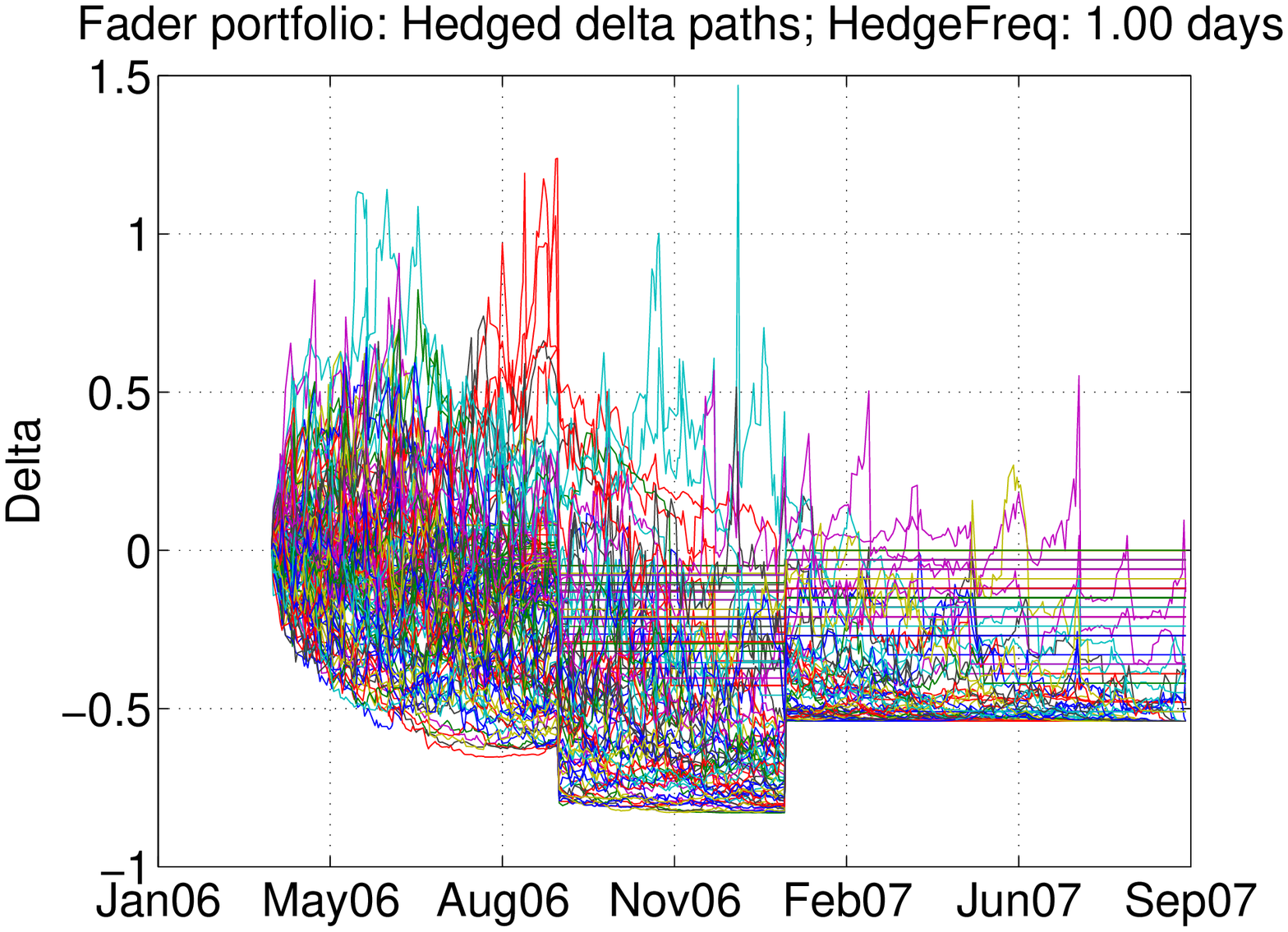}
	\includegraphics[width=0.45\textwidth]{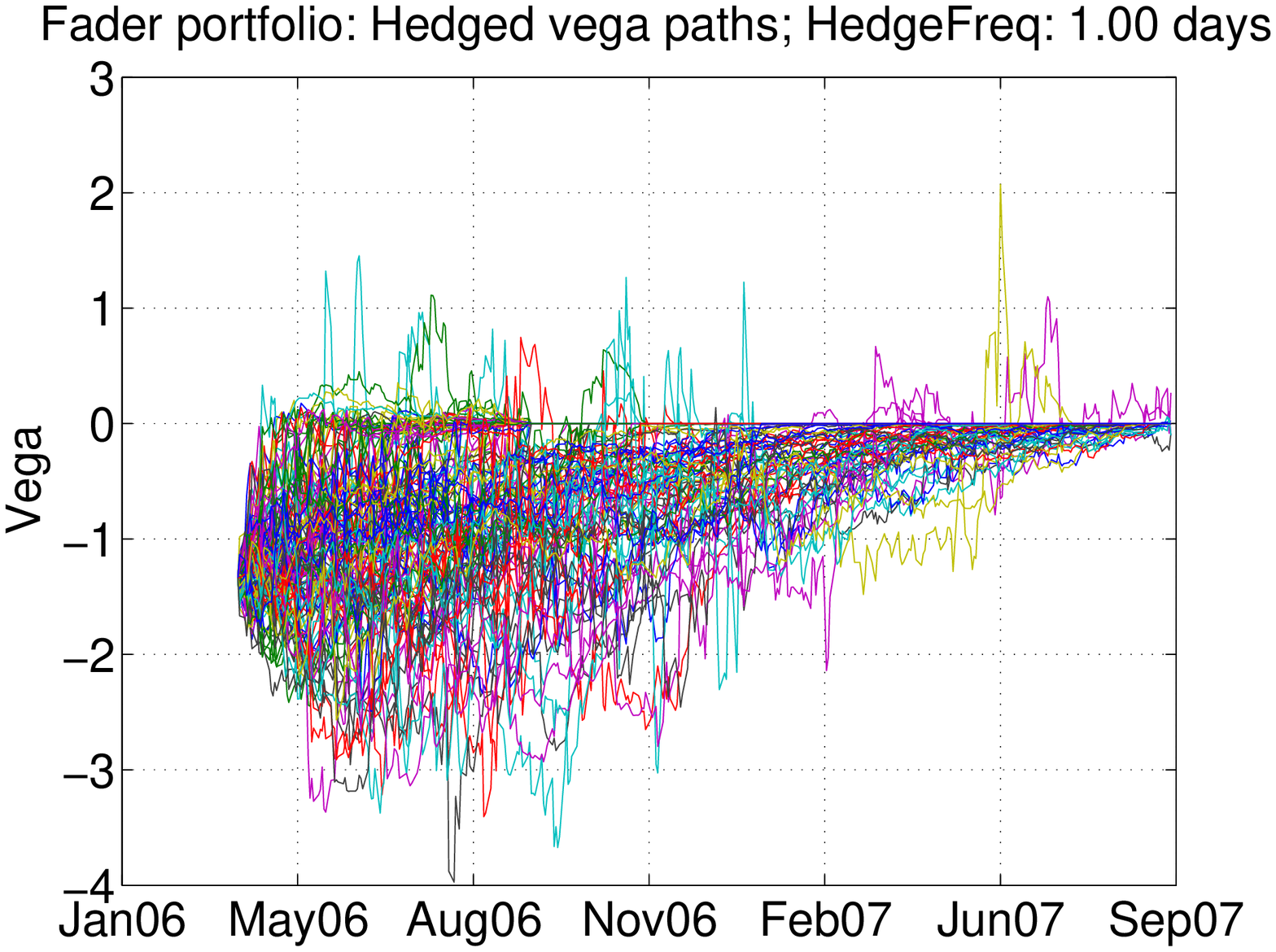}
	\caption{Delta (left) and vega (right) paths for a portfolio of forward faders valued with volga-vanna.}
	\label{fig:Fader_delta_vega}
\end{figure}

Figure \ref{fig:Fader_delta_vega} shows the delta paths on the left hand side and $\vartheta$ paths on the right hand side (the latter are labelled as ``Vega'') of the portfolio of fader options. The delta paths show three clear periods of time which correspond to the different expiry dates of the components of the portfolio. The first period corresponds to a portfolio with the three deals until the expiry of the fader with 17\% notional after 0.5 years. The second period goes from 0.5 to 0.8 years with only two deals until the expiration of the fader with 29\% notional. The last period goes up to the maturity of the fader with 54\% notional (1.5 years). Some horizontal straight lines from a given point of time may appear. They occur when the portfolio of options does not add any more notional because the barrier level has been breached and the fader with 17\% (and no barriers) has already expired. These straight lines appear because a forward contract with fixed notional has constant delta, no vega and a perfect hedge (hedging costs are zero). The left plot of figure \ref{fig:Fader_delta_vega} presents that deltas get confined. The lower limit in the first period is -0.66 (the sum of the deltas of the three plain forward deals: 17\% -54\% - 29\% = 66\%). The second period has a minimum delta of -0.83 (the sum of deltas of the living two deals: 54\% + 29\% = 83\%). Finally the last period has a lower delta threshold of -0.54, corresponding to the only living deal with 54\% notional. The upside of delta will depend on the position of the spot with respect to the barriers. The sensitivity $\vartheta$ (right plot of figure \ref{fig:Fader_delta_vega}) is rather confined (see that for the double-no-touch it could get up to 200). Therefore, model risk for this portfolio should be intuitively much lower than for the double-no-touch option of section \ref{sec:DNT}.

\begin{figure}[htbp]
	\centering
	\includegraphics[width=0.45\textwidth]{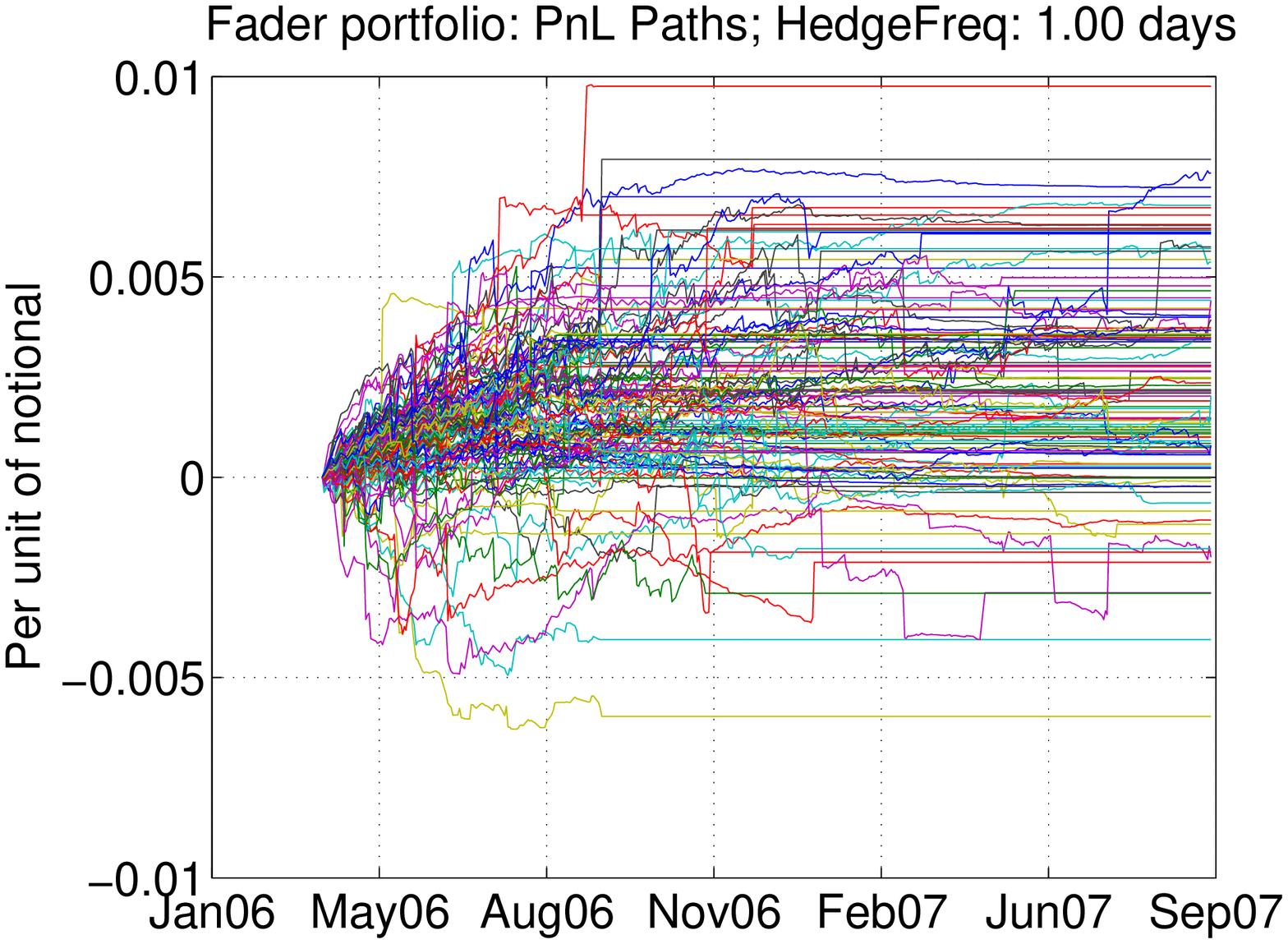}
	\includegraphics[width=0.45\textwidth]{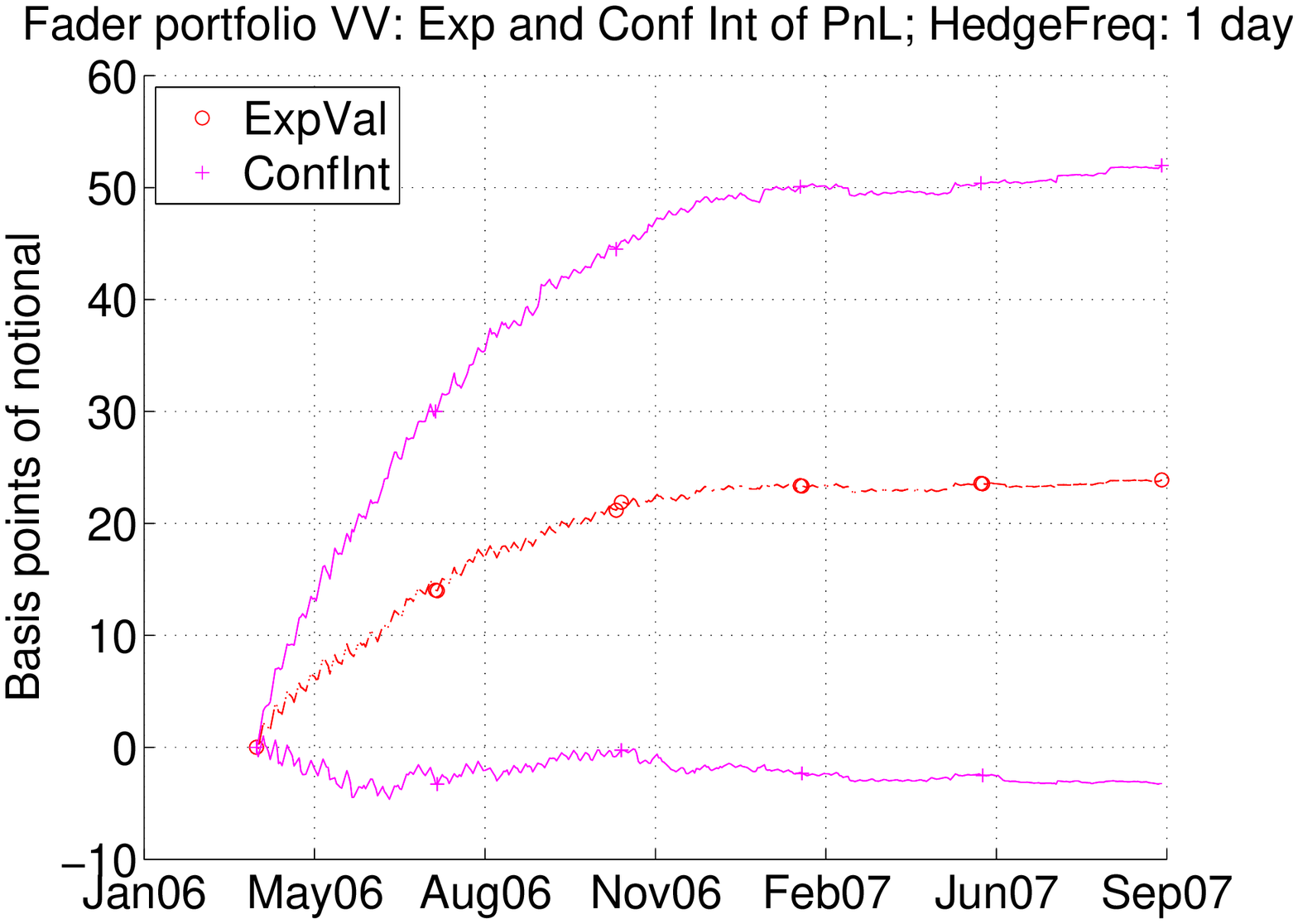}
	\caption{Comparison of the expected value (left) and standard deviation (right) of P\&L for volga-vanna (``VV'') and Black Scholes (``BS'') models.}
	\label{fig:Fader_ExpVal_StdDev}
\end{figure}

Figure \ref{fig:Fader_ExpVal_StdDev} shows the paths with the evolution of the hedging earnings per unit of notional, $\Pi^{Tot}_{t}$ (left plot), and the evolution of their expected value the two confidence intervals (defined in the same way as in section \ref{sec:DNT}) with a degree of significance of 35\% (right plot). The straight lines of the left plot of figure \ref{fig:Fader_ExpVal_StdDev} of the hedging earnings correspond to situations in which the knock-out barriers have already been touched and the notional is constant. In this situation, the portfolio is equivalent to a forward contract with constant delta and an exact replication portfolio (the hedging earnings keep constant). The right plot of figure \ref{fig:Fader_ExpVal_StdDev} shows that the expected earnings are around 22 basis points whereas the uncertainty (difference between the expected value and expected shortfall) is very small (approximately 25 basis points). The relative measure of model risk defined in section \ref{sec:ModelRiskMeasure} or the expected shortfall is around 3 basis points. This means that the expected earnings would compensate the uncertainty of the hedging cost and the initial premium assigned to this portfolio would be just fair under the market assumptions.

\section{Conclusions}
\label{sec:Conclusions}

This paper introduces a relative model risk measure of a given product priced with a particular model with respect to a reference model. It is based on the simulation of the hedging strategy of the product priced with its model assuming that the market is driven by the reference model (e.g. Heston model) calibrated to real market data. The relative model risk measure is calculated as the expected shortfall of the hedging cost for a given significance level. This level allows setting the risk aversion. This model risk measure can be directly interpreted as the provision or amount of money to reserve in order to be safe from a risk management point of view. 

This methodology has been applied to foreign exchange double-no-touch options priced and managed with volga-vanna and Black Scholes models. It has been verified that the market price is approximately equal to the model price plus the expected hedging loss for both models (Black Scholes and volga-vanna). The method has concluded that the volga-vanna model shows around half the model risk compared to Black Scholes for a given deal. However, volga-vanna model still shows a remarkable amount of model risk for double-no-touch options. The same volga-vanna model has been thereafter used to value a portfolio of forward faders and the method has concluded that model risk is irrelevant (same model may involve different model risk depending on the product).

This methodology can also be applied to model validation. Model implementation errors or inappropriate market hypothesis will show up in the measure of model risk. In addition, jumps or estrange behaviour of the hedging earnings, hedge ratios or premium can be easily identified for particular paths. This would allow identifying potential problems and understanding very well the risks of the product under consideration.

When a deal is closed, the average hedging loss can be easily estimated through fair value adjustment (FVA) by comparing the price of the model used to manage the position (the model which gives the hedging ratios) with another better pricing model (perhaps too slow for daily management) which reflects better the hypothesis of the market. This provision needs to be calculated and accounted only once, when the operation is closed. However, the uncertainty of the hedging loss is considerably more difficult to estimate and may have a bigger impact.

\begin{table}[htbp]
  \centering
    \begin{tabular}{|p{13cm}|}
    \hline
    \textbf{Ackknowledgement}: The authors want to thank the discussions with Servando Arbol\'{i} and Francisco Amaya and their contribution to the improvement of the paper. \\
    \hline
    \end{tabular}
  \label{tab:Acknowledgement}
\end{table}

%


\begin{thebibliography}{30}

\bibitem{Artzner1999}
Artzner P., Delbaen F., Eber J. M., Heath D., ``Coherent Measures of Risk'', \emph{Mathematical Finance}, Vol. 9, No. 3, pp. 203-228, July 1999.

\bibitem{Black1973}
Black F., Scholes M., ``The pricing of options and corporate liabilities'', \emph{Journal of Political Economy}, Vol. 81, No. 3, pp. 637-654, 1973.

\bibitem{Carr2002}
Carr P., Madan D., ``Towards a Theory of Volatility Trading', \emph{Volatility}, Risk publications, Robert Jarrow, ed., pp. 417-427. 2002. Available at: http://www.math.nyu.edu/research/carrp/research.

\bibitem{Carr2007} Carr P., Sun J., ``A new approach for option pricing under stochastic volatility', \emph{Review of Derivatives Research}, Vol. 10, No. 2, 87-150, May 2007. Available at: http://www.math.nyu.edu/research/carrp/research.

\bibitem{Castagna2007}
Castagna A., Mercurio F., ``The vanna-volga method for implied volatilities'', \emph{Risk Magazine}, March 2007.

\bibitem{Castagna2007b}
Castagna A., Mercurio F., ``Consistent pricing of FX options'', \emph{Internal Report}, Banca IMI, 2005, available at ``www.fabiomercurio.it/consistentfxsmile2b.pdf''.

\bibitem{Cont2006}
Cont R., ``Model uncertainty and its impact on the pricing of derivative instruments'', \emph{Mathematical Finance}, Vol. 16, No. 3, pp. 519-547, July 2006.

\bibitem{Corielli2006}
Corielli F., ``Hedging with energy.'', \emph{Mathematical Finance}, Vol. 16, No. 3, pp. 495-517, July 2006.

\bibitem{Dupire1994}
Dupire B., ``Pricing with a smile'', \emph{Risk Magazine}, pp. 18-20, 1994.

\bibitem{Elices2007}
Elices A., ``Models with time-dependent parameters using transform methods: application to Heston's model'', \emph{arXiv:0708.2020v2 [q-fin.PR]}, October 2010. Available at ``http://arxiv.org/abs/0708.2020''.

\bibitem{Elices2009}
Elices A., ``Affine concatenation'', \emph{Wilmott Journal}, Vol. 1, No. 3, pp. 155-162, June 2009.

\bibitem{Gatheral2006}
Gatheral J., ``The volatility surface: a practicioner's guide'', \emph{John Wiley and Sons, Inc}, 2006.

\bibitem{Heston1993}
Heston S. L., ``A closed-form solution for options with stochastic volatility with applications to bond and currency options.'', \emph{The Review of Financial Studies}, No. 6, pp. 327-343, 1993.

\bibitem{Kurpiel1998}
Kurpiel A., Roncalli T., ``Option hedging with stochastic volatility'', \emph{Social Science Research Network}, December 2006. Available at: http://ssrn.com/abstract=1031927.

\bibitem{Lipton2002}
Lipton A., McGhee W., ``Universal Barriers'', \emph{Risk Magazine}, published on line, May 2002.

\bibitem{Merton1993}
Merton R., ``Option pricing when underlying stock returns are discontinuous.'', \emph{Journal of Financial Economics}, Vol. 3, No. 1-2, pp. 125-144, 1976.

\bibitem{Ren2007}
Ren Y., Madan D., Quian M., ``Calibrating and pricing with embedded local volatility models'', \emph{Risk Magazine}, September 2007.

\bibitem{Schoutens2003}
Schoutens W., Simons E., Tistaert J., ``A Perfect Calibration! Now what?'', \emph{Wilmott Magazine}, March 2004.

\bibitem{Schweizer1992}
Schweizer M., ``Mean-variance hedging for general claims.'', \emph{Annals of Applied Probability}, Vol. 2, pp. 171-179, 1992.

\bibitem{Sharpe1964}
Sharpe W., ``Capital Asset Prices: A theory of market equilibrium under conditions of risk.'', \emph{The Journal of Finance}, Vol. 19, No. 3, pp. 425-442, September 1964.

\bibitem{Wystup2003}
Wystup U., ``The market price of one-touch options in foreign exchange markets'', \emph{Derivatives week}, Vol. XII, No. 13, London 2003.

\bibitem{Zhou2007}
Jin H., Zhou X., ``Continuous-Time Markowitz's problems in an incomplete market, with no-shorting portfolios.'', \emph{Stochastic Analysis and Applications}, Vol. 2, pp. 435-459, 2007.

\end{thebibliography}
\end{document}